\documentclass[aps,prb,preprint,superscriptaddress]{revtex4-2}

\usepackage{siunitx}
\usepackage{physics}
\newcommand{\FZMO}{\ce{(Fe_{0.95}Zn_{0.05})2Mo3O8}}
\newcommand{\DeltaSM}{\Delta S_{\mathrm{M}}}
\newcommand{\DeltaSE}{\Delta S_{\mathrm{E}}}
\newcommand{\DeltaSME}{\Delta S_{\mathrm{MF}}}

\newcommand{\DeltaTad}{\Delta T_{\mathrm{ad}}}
\usepackage{chemformula}
\usepackage{multirow}
\let\ce\ch
\usepackage{color}
\usepackage{ulem}

\begin{document}

\title{Magnetoelectric coupling and its impact on the multicaloric effect}

\author{Kentaro Ino}
\affiliation{Department of Physics, Institute of Science Tokyo, Meguro 152-8551, Japan}

\author{Keisuke Matsuura}
\email[]{matsuura.k.644d@m.isct.ac.jp}
\affiliation{Department of Physics, Institute of Science Tokyo, Meguro 152-8551, Japan}

\author{Tetsuya Nomoto}
\affiliation{RIKEN Center for Emergent Matter Science, Wako 351-0198, Japan}

\author{Takashi Kurumaji}
\affiliation{Division of Physics, Mathematics and Astronomy,California Institute of Technology, Pasadena, California 91125, USA}

\author{Yoshinori Tokura}
\affiliation{RIKEN Center for Emergent Matter Science, Wako 351-0198, Japan}
\affiliation{Tokyo College and Department of Applied Physics, University of Tokyo, Tokyo 113-8656, Japan}

\author{Fumitaka Kagawa}
\affiliation{Department of Physics, Institute of Science Tokyo, Meguro 152-8551, Japan}
\affiliation{RIKEN Center for Emergent Matter Science, Wako 351-0198, Japan}

\date{\today}

\begin{abstract}
The multicaloric effect, which represents the reversible entropy change that occurs when both external magnetic and electric fields are applied, is an interesting phenomenon characteristic to multiferroics. Targeting the multicaloric effect in the typical multiferroic \ce{(Fe_{0.95}Zn_{0.05})2Mo3O8}, we experimentally evaluate the isothermal entropy change due to a magnetoelectric cross-correlation. A pronounced cross-correlation-derived isothermal entropy change is found at the magnetic ordering temperature. By exploring the literature, we suggest that a magnetic phase transition 
 may tend to involve a considerably larger cross-correlation-derived entropy change than that derived from the linear magnetoelectric effect.
\end{abstract}

\maketitle

\section{Introduction}
The mutual coupling of electricity and magnetism has been extensively studied since the discovery of the giant magnetoelectric effect in multiferroics \cite{kimura2003magnetic}.
Owing to the magnetoelectric cross-correlation, a magnetic phase transition can be accompanied by a change in electric polarization $P$ and $P$ (or magnetization $M$) can be induced by a magnetic field $B$ (or an electric field $E$). Such cross-correlational electromagnetic responses have been a central issue since the early stage of research on magnetoelectric multiferroics \cite{tokura2014multiferroics, fiebig2016evolution}.
Recently, magnetoelectric multiferroics have also been found to be an intriguing platform for realizing thermal-related phenomena, such as the thermal Hall effect \cite{ideue2017giant,kim2024thermal}, thermal-diode effect \cite{hirokane2020nonreciprocal}, spin Seebeck effect \cite{takagi2016thermal,seki2015thermal}, and magnetoelectric-coupling-based caloric effect \cite{ikeda2023magnetoelectrocaloric}.
These advances evoke fundamental interest in how the coupled magnetic and electric degrees of freedom in magnetoelectric multiferroics are involved in thermal-related phenomena. Here, we study this issue by focusing on the caloric effect, which is one of the most fundamental thermal-related phenomena.

The caloric effect denotes the reversible change in entropy under an isothermal process or temperature under an adiabatic process that occurs when an external field is applied to a material.
In an isothermal process, the temperature of the material is kept constant and the application of an external field leads to the entropy change, called the isothermal entropy change.
In an adiabatic process, on the contrary, the entropy of the material is kept constant, and the application of an external field leads to the temperature change, called the adiabatic temperature change.
Hence, the performance of the caloric effect is normally evaluated by the isothermal entropy change $\Delta S_{\rm iso}$ or  
the adiabatic temperature change $\DeltaTad$ \cite{moya2014caloric}.
An approximate relation, $\DeltaTad \simeq -(T/C_{p})\Delta S_{\rm iso}$, holds between $\DeltaTad$ and $\Delta S_{\rm iso}$ as long as the specific heat $C_p$ does not vary significantly.
The conventional caloric effect by a single external field
has been referred to as the monocaloric effect \cite{hou2022materials} and, in this definition, magnetocaloric \cite{tishin2016magnetocaloric}, electrocaloric \cite{liu2016direct}, elastocaloric \cite{cong2019colossal}, and barocaloric effects \cite{matsunami2015giant} under swept magnetic field, electric field, strain field, and pressure, respectively, are all classified as the monocaloric effect.

Recently, the multicaloric effect, which was originally defined as the caloric effect induced by simultaneous sweeping of two or more external fields, has been attracting attention \cite{starkov2014multicaloric,starkov2016generalized,hou2022materials,stern2018multicaloric}.
The isothermal entropy change in the multicaloric effect is expected to differ from the simple sum of the relevant monocaloric effects and to include nontrivial additional contributions unique to the multicaloric effect. In this context, note that the meaning of the term ``multicaloric effect'' varies in the literature and sometimes refers to the caloric effect exhibited by systems with multiple orders, regardless of the number of external fields. To avoid this confusion, we introduce the new term “multifield caloric effect” (MFCE), which denotes the caloric effect induced by sweeping of multiple external fields (for details, see  Sec.~II). 
The MFCE has been experimentally studied in materials such as \ce{BiCu3Cr4O12} \cite{kosugi2021giant}, Ni-Mn-Ga-In alloys \cite{qian2022multicaloric}, Ni-Mn-Ga-Cu alloys \cite{manosa2023cross}, \ce{Fe49Rh51} \cite{manosa2023cross}, and Fe-Rh/\ce{BaTiO3} \cite{liu2016large}, but the additional entropy change accompanying the MFCE has not been clearly discussed except for in Ref.~\cite{manosa2023cross}. 
In particular, to our knowledge, the MFCE in magnetoelectric materials has not been reported thus far. 
The purpose of this study is to reveal the fundamental issue of how the magnetoelectric cross-correlation in multiferroics can contribute to the MFCE.

The structure of this paper is as follows.
In Sec.~II, we provide a thermodynamic framework of the MFCE, with particular attention 
to clarifying previous ambiguous terminology of multicaloric effect.
These ambiguities are resolved through a classification based on the number of order parameters and applied fields.
Section~III introduces the target material of this study, $\FZMO$. 
Section~IV details the experimental methods.
Section~V presents the experimental results concerning the additional isothermal entropy change  that is unique to the MFCE, which we abbreviate as the ``multifield contribution''.
In Sec.~VI, we applied our method to various magnetoelectric multiferroics to identify the underlying mechanisms of the multifield contribution.
We observe that materials showing magnetic phase transition tend to involve a pronounced multifield contribution, whose magnitude is one or two orders of magnitude greater than that of the contribution in linear magnetoelectric materials.
Finally, Sec.~VII summarizes the findings and provides concluding remarks.

\section{Thermodynamics of the multifield caloric effect}
\begin{figure}[t]
\centering
\includegraphics[width =12cm]{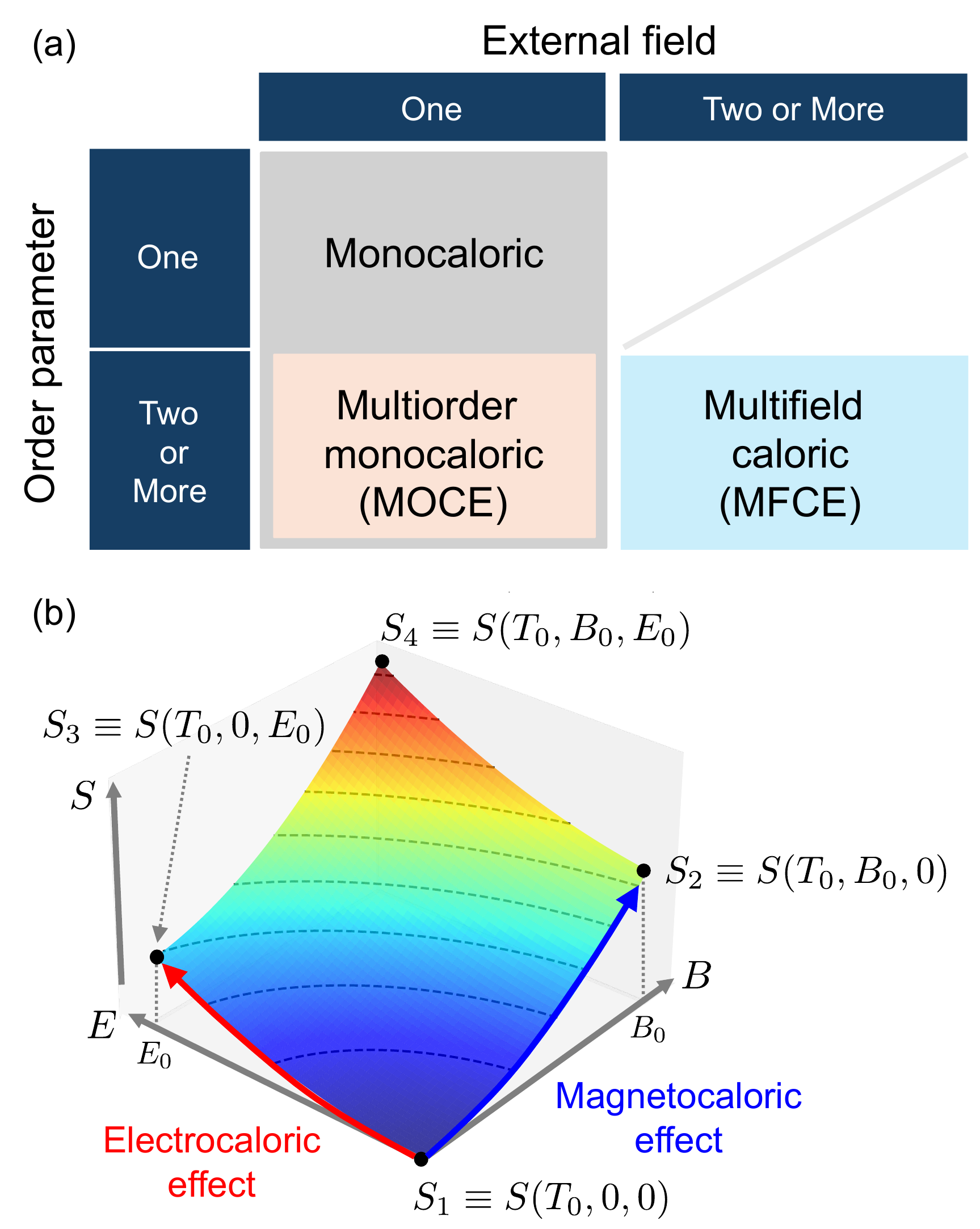}
\caption{
(a) Definitions of the monocaloric, multiorder monocaloric (MOCE), and multifield caloric effects (MFCE).
(b) Schematic landscape of the entropy $S = S(T_0,B,E)$ at temperature $T_0$. The isothermal entropy change under sweeping of both electric and magnetic fields is represented by the difference in entropy between $S_1$ and $S_4$: $\Delta S_{\rm iso} = S_4 - S_1$.
}
\label{fig3}
\end{figure}

The multicaloric effect was originally defined as the adiabatic temperature change or the isothermal entropy change that occurs when two or more external fields are applied to a material \cite{qian2022multicaloric,starkov2014multicaloric,stern2018multicaloric}.
However, when a change in two or more order parameters is relevant to the caloric effect, the term ``multicaloric effect'' is sometimes used even for cases when a single external field is applied \cite{vopson2012multicaloric,vopson2013theory,VOPSON201614,ikeda2023magnetoelectrocaloric}.
Because the meaning of the multicaloric effect varies in the literature, we try to avoid using the term ``multicaloric effect''.
Instead, we introduce new terms, (i) the multifield caloric effect (MFCE) and (ii) the multiorder monocaloric effect (MOCE) (see Fig.~\ref{fig3}(a)), as well as the number of applied external fields and the number of involved order parameters (or, more simply, extensive valuables) and to clarify what ``multi'' means.
The MFCE refers to the caloric effect when multiple external fields are simultaneously applied, whereas the MOCE refers to the caloric effect when two or more order parameters vary when a single external field is applied.
The MOCE in our study has the same meaning as ``the cross-caloric effect'' in Refs.~\cite{edstrom2020prediction,planes2014thermodynamics,li2024cross} or ``the magnetoelectric-coupling-based caloric effect'' in Refs.~\cite{ikeda2023magnetoelectrocaloric}.
Note that when more than one external field is used in caloric experiments, multiple extensive conjugate valuables are inevitably involved in the response.
In Fig.~\ref{fig3}(a), we therefore exclude the situation in which only one order parameter changes under multiple fields.

In the MFCE, the resulting isothermal entropy change is not represented by the sum of the relevant monocaloric effects; thus it involves an additional term, namely, the multifield contribution. In the following, we provide a thermodynamic definition of the multifield contribution to the MFCE, which is the focus of this study, and show that it is related to the magnetoelectric cross-correlation.
For thermodynamic arguments of its relation to the MFCE and MOCE, see Appendix~A.

Here, we consider the situation in which external magnetic and electric fields are simultaneously applied and consider the isothermal entropy change by referring to the schematic entropy landscape with respect to the electric and magnetic field plane (Fig.~\ref{fig3}(b)).
For clarity, we define four coordinate points: $S_1 \equiv S(T_0,0,0)$, $S_2 \equiv S(T_0, B_0,0)$, $S_3 \equiv S(T_0,0,E_0)$, and $S_4 \equiv S(T_0,B_0,E_0)$.
The entropy $S(T, B, E)$ is a state quantity that is uniquely determined at equilibrium, except for at a first-order phase transition point.
The isothermal entropy change $\Delta S_{\mathrm{iso}}$ under the simultaneous application of electric and magnetic fields is represented by the difference in entropy between points $S_1$ and $S_4$ in Fig.~\ref{fig3}(b):
\begin{equation}
\label{eq5}
    \begin{aligned}
      \Delta S_{\mathrm{iso}} =& S_{4} - S_{1} \\
      =& (S_{2} - S_{1}) + (S_{3} -S_{1}) + (S_4 - S_3 - S_2 + S_1)\\
      =& \DeltaSM + \DeltaSE + \DeltaSME,
    \end{aligned}
\end{equation}
where $\DeltaSM \equiv S_2 - S_1$ corresponds to the magnetocaloric effect at $E=0$, $\DeltaSE \equiv S_3 - S_1$ represents the electrocaloric effect at $B=0$, and $\DeltaSME \equiv S_4 - S_3 - S_2 + S_1$ is an additional term that is unique to the MFCE.
Equation~(\ref{eq5}) explicitly states that $\Delta S_{\rm iso}$ is not represented by the sum of the two relevant monocaloric effects. Note that $\DeltaSME$ becomes more or less finite in any material because the simultaneous application of $B$ and $E$ inevitably forces breaking of both the space-inversion and time-reversal symmetries, activating magnetoelectric coupling in a broad sense. 
However, inherent and strong magnetoelectric coupling is characteristic of magnetoelectric multiferroic materials and thus a considerable $\DeltaSME$ is expected, which is the focus of the present study.

\section{Introduction to $\FZMO$}
To examine the multifield contribution $\DeltaSME$ discussed here, we choose $\FZMO$ for the following two reasons.
First, it shows two kinds of magnetic order in the low-magnetic-field region below $\SI{9}{T}$, accompanied by electric polarization arising from magnetoelectric responses \cite{kurumaji2015doping}.
Second, the first-order phase transition field is moderate ($\sim \SI{3}{T}$; see Fig.~\ref{fig1}(b)), and it is easily accessible by experiments.

The parent Fe$_2$Mo$_3$O$_8$ is a polar crystal that belongs to the space group $P6_3mc$ \cite{mccarroll1957some} (Fig.~\ref{fig1}(a)), allowing crystallographic polarization along the $c$ axis. 
Whereas Mo$^{4+}$ ions forming spin-triplet trimers do not contribute to the magnetism \cite{cotton1964metal}, two types of Fe$^{2+}$ ions with different oxygen coordinations (labeled A and B in Fig.~\ref{fig1}(a)) dominate the magnetism of this compound. Nonmagnetic Zn$^{2+}$ ions preferentially substitute the A-site Fe$^{2+}$ ions \cite{FZMO_prefer}.
The magnitudes of the magnetic moments at the A site and B site are different \cite{bertrand1975structural,reschke2020structure}.
The magnetic ground state in the low-magnetic-field region is an antiferromagnetic (AFM) phase and a ferrimagnetic (FRI) phase emerges under high magnetic fields (Fig.~\ref{fig1}(b)) \cite{matsuura2023thermodynamic}.
Both the linear magnetoelectric effect $(\Delta P\propto B)$ and second-order magnetoelectric effect $(\Delta P\propto B^2)$ are allowed in the FRI phase with the magnetic point group $6m'm'$ \cite{wang2015unveiling,kurumaji2015doping}, whereas only the second-order magnetoelectric effect is allowed in the AFM phase with magnetic point group $6'mm'$.

\section{Experimental details}
The $\FZMO$ single crystals used for this study were grown by the chemical vapor transport method \cite{kurumaji2015doping}.
We used two samples labeled A and B for this study. 
Sample A ($\SI{4.2}{mm^2}\times \SI{650}{\mu m}$) was used for the magnetocaloric and electric polarization experiments, and sample B ($\SI{2.4}{mm^2}\times \SI{150}{\mu m}$) was used for $P$-$E$ curve measurements.

\begin{figure*}[tbp]
    \centering
    \includegraphics[width = \hsize]{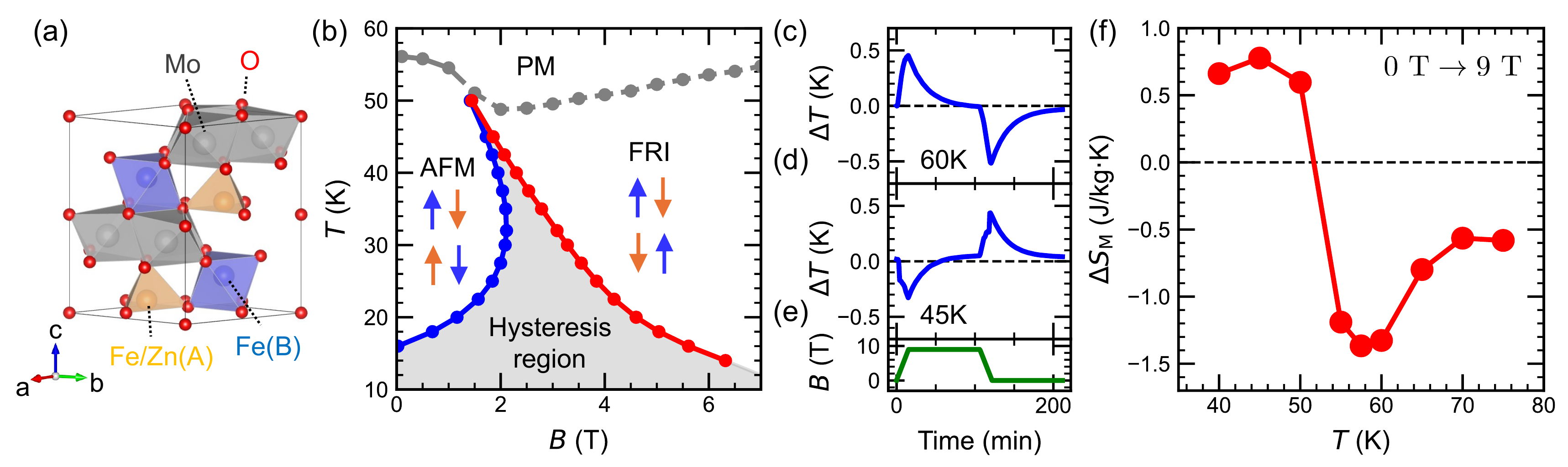}
    \caption{
     (a) Crystal structure of \ce{(Fe_{0.95}Zn_{0.05})2Mo3O8}. There are two inequivalent Fe sites (labeled A and B).
    (b) Phase diagram of \ce{(Fe_{0.95}Zn_{0.05})2Mo3O8}, reproduced from a previous report \cite{matsuura2023thermodynamic}.
    The transition fields from the AFM to FRI phases and from the FRI to the AFM phases are indicated by red and blue filled circles, respectively.
    The gray filled circles represent the PM-AFM or PM-FRI boundary.
    A hysteresis region between the AFM and FRI phases is shown as a gray hatched area. 
    Orange and blue arrows in the schematic represent spins at Fe(A) and Fe(B) sites, respectively.
    (c), (d) Time profiles of the sample temperature change $\Delta T = T(t) - T(0)$ under adiabatic conditions at $\SI{60}{K}$ and $\SI{45}{K}$.
    (e) Time profile of the applied magnetic field in the adiabatic temperature change measurement.
    (f) Temperature dependence of the isothermal entropy change $\DeltaSM$ under a magnetic field sweep of $\SI{0}{T} \rightarrow \SI{9}{T}$.
    During the field sweep, the first-order phase transition occurs below $\SI{50}{K}$, and the PM phase appears above $\SI{56}{K}$. 
    }
    \label{fig1}
\end{figure*}

For the magnetocaloric measurement, the adiabatic temperature measurement was performed with a physical property measurement system (PPMS, Quantum Design). High vacuum mode was selected to achieve high vacuum conditions ($\sim$ a few mPa) in the sample space.
The adiabatic temperature change was measured by our own setup on the PPMS resistivity puck. Our setup consisted of two Pt thermometers; one Pt thermometer was hung by Pt‒W wires ($\phi = \SI{40}{\mu m}$) to monitor the sample temperature, and the other thermometer was placed in the stage area of the PPMS resistivity puck to monitor the puck temperature. The sample was attached to the former thermometer with Apiezon N grease. While sweeping magnetic fields to $\SI{9}{T}$ along the $c$ axis, we measured the temperatures using both Pt thermometers. Raw temperature data are often affected by drift of the base temperature and thermal diffusion from wires \cite{yonezawa2013first}; the base-temperature drift has been corrected in Figs.~\ref{fig1}(c) and \ref{fig1}(d), and the heat dissipation through the wires has been analytically removed in Fig.~\ref{fig1}(f) (for details, see Appendix~B).

For the electric polarization measurement, a commercial cryostat equipped with a superconducting magnet was used for temperature and magnetic field control. The sample $c$ plane surfaces were covered with Ag paste to make capacitors. The pyroelectric current in the $c$ direction was collected by an electrometer (Keithley, 6517B) while the temperature or magnetic field was swept. The electric polarization was obtained by integrating the measured pyroelectric current with respect to time. The temperature sweep rate was $\SI{0.05}{K/sec}$ ($\SI{3}{K/min}$), and the magnetic field sweep rate was $\SI{0.01}{T/sec}$ ($\SI{100}{Oe/sec}$).
Before each magnetic field scan, the sample temperature was increased to $\SI{100}{K}$ to eliminate the effect of phase coexistence of the AFM and FRI phases.
For $P$-$E$ curve measurements, we used a ferroelectric tester (Radiant, Precision Premier II) and measured the electric polarization up to $\SI{200}{V}$ in standard bipolar mode. The drive frequency was $\SI{200}{Hz}$. 

\section{Results}
\subsection{Magnetocaloric effect}
As the basic caloric property of this compound, we show the magnetocaloric effect, which denotes the adiabatic temperature change or the isothermal entropy change that occurs under a magnetic field sweep, in \ce{(Fe_{0.95}Zn_{0.05})2Mo3O8}.
The phase diagram of \ce{(Fe_{0.95}Zn_{0.05})2Mo3O8} is shown in Fig.~\ref{fig1}(b).
When the magnetic field is applied along the $c$ axis, the first-order phase transition between the AFM and FRI phases occurs below $\SI{50}{K}$.
Figures~\ref{fig1}(c) and \ref{fig1}(d) show the time evolution of the sample temperature change $\Delta T = T(t) - T(0)$ at $\SI{60}{K}$ and $\SI{45}{K}$ during magnetic field sweeping under a nominally adiabatic environment, $B = \SI{0}{T} \rightarrow \SI{9}{T} \rightarrow \SI{0}{T}$, as shown in Fig.~\ref{fig1}(e).
At $\SI{60}{K}$, the sample temperature increases ($\Delta T >0$) in the field-applying process and decreases ($\Delta T <0$) in the field-removing process (Fig.~\ref{fig1}(c)), which is typical behavior of the magnetocaloric effect in a paramagnetic (PM) phase. This standard magnetocaloric effect indicates a decrease in the entropy with the application of a magnetic field, which results in an increase in the sample temperature under adiabatic conditions (see also Fig.~\ref{fig1}(f)).
In contrast, at $\SI{45}{K}$, the application of a magnetic field results in a negative sample temperature change $\Delta T <0$ (Fig.~\ref{fig1}(d)). 
This phenomenon is known as the inverse magnetocaloric effect \cite{krenke2005inverse,tran2022direct}, which occurs when the application of a magnetic field results in an increase in the entropy.
As reported in the previous study, the entropy of this compound increases when the magnetic field is applied along the $c$ axis below $\SI{45}{K}$ \cite{matsuura2023thermodynamic},
which is consistent with a negative sample temperature change ($\Delta T < 0$) in the field-applying process. 
Figure \ref{fig1}(f) shows the temperature dependence of the isothermal entropy change in the magnetocaloric effect,  $\DeltaSM$, which is calculated from $\DeltaSM = -(C_{p}/T)\DeltaTad$.
The specific heat $C_p$ is taken from Ref.~\cite{matsuura2023thermodynamic}.
We experimentally obtained $\DeltaTad$ after taking into account the thermal diffusion through wires from the sample temperature change $\Delta T$ (for details, see Appendix~B).
The result clearly indicates that the magnetic phase transition between the AFM and FRI phases causes the inverse magnetocaloric effect. 
These observations corroborate that the thermal response of magnetoelectric \ce{(Fe_{0.95}Zn_{0.05})2Mo3O8} is closely linked to its magnetism. 

Note again that the magnetocaloric effect in $\FZMO$ is classified as the MOCE in our definition, because the magnetic field affects both the magnetization and the polarization.
The magnetoelectric cross-correlation must therefore be involved in the isothermal entropy change $\DeltaSM$ shown in Fig.~\ref{fig1}(f). However, this contribution is experimentally indistinguishable as long as a standard magnetocaloric experiment is performed (for details, see Appendix~A2 and Eq.~(\ref{last_result})). 
Accordingly, in the present study, we focus on the MFCE to clarify the contribution from the magnetoelectric cross-correlation.

\subsection{Multifield contribution}
\begin{figure*}[t]
    \centering
    \includegraphics[width = \hsize]{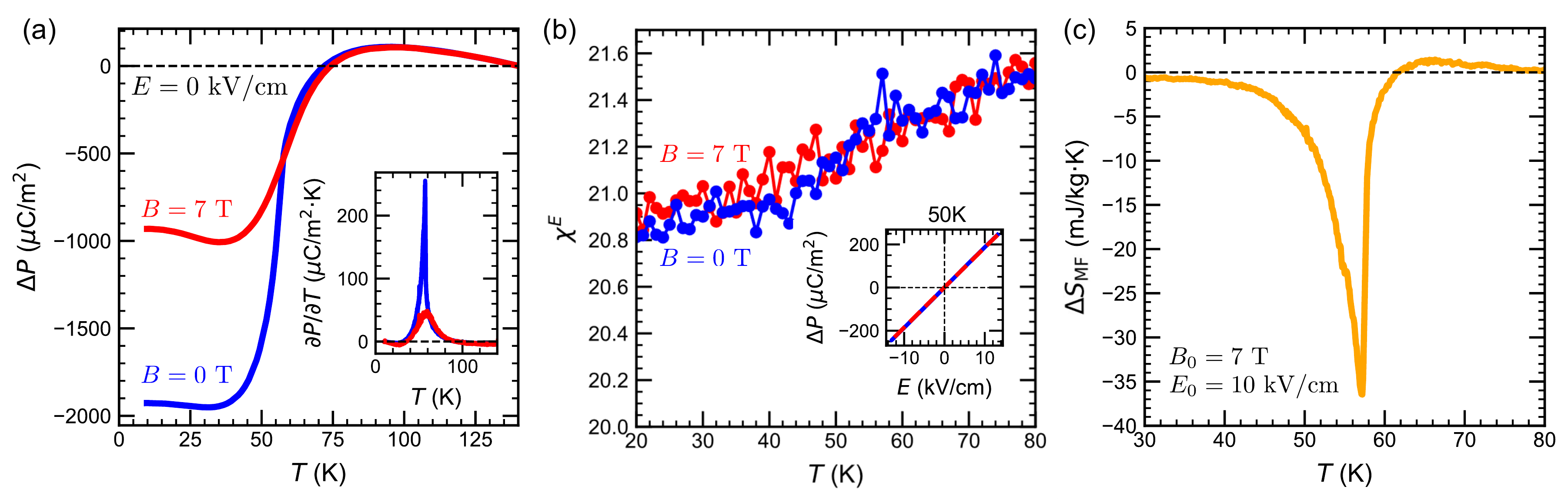}
    \caption{
    (a) Temperature dependence of the polarization $\Delta P = P_{E=0}(T,B) - P_{E=0}(T_0,B)$ at $B=\SI{0}{T}$ (blue) and $\SI{7}{T}$ (red) in a cooling process with reference to $T_0=\SI{140}{K}$ since the present compound is a polar crystal and hosts finite polarization even without magnetic order.
    The inset shows the temperature dependence of $\partial P_{E=0}/\partial T$ at $B=\SI{0}{T}$ (blue) and $\SI{7}{T}$ (red).
    (b) Temperature dependence of the electric susceptibility $\chi^E(T,B)$ at $B=\SI{0}{T}$ (blue) and $\SI{7}{T}$ (red) obtained from the $P$-$E$ curve measurements shown in the inset. The inset shows the electric-field-induced change in the polarization: $\Delta P =P(T,B,E) - P(T,B,0)$ under $B=\SI{0}{T}$ (blue) and $\SI{7}{T}$ (red) at $T=\SI{50}{K}$.
    (c) Temperature dependence of the multifield contribution to the MFCE $\DeltaSME$ at $B_0 = \SI{7}{T}$ and $E_0 = \SI{10}{kV/cm}$.
    }
    \label{fig4}
\end{figure*}

We proceed to evaluate the multifield contribution to the MFCE, $\DeltaSME$, in \ce{(Fe_{0.95}Zn_{0.05})2Mo3O8} with $B_0 = \SI{7}{T}$ and $E_0 = \SI{10}{kV/cm}$.
We first calculate $\DeltaSME$ by using the Maxwell relation:
\begin{equation}
    \begin{aligned}
        \DeltaSME &= S_4 - S_3 - S_2 + S_1  = (S_4-S_2) - (S_3 - S_1)\\
      & = \int_0^{E_0}\left[\left(\pdv{S}{E}\right)_{T,B=B_0}-\left(\pdv{S}{E}\right)_{T,B=0}\right]\dd{E} \\ 
      & = \int_0^{E_0}\left[\left(\pdv{P}{T}\right)_{B=B_0,E}-\left(\pdv{P}{T}\right)_{B=0,E}\right]\dd{E}. \\
    \end{aligned}
\end{equation}
We measure the $P$-$E$ curve at $B=\SI{0}{T}$ (the AFM phase) and $\SI{7}{T}$ (the FRI phase) in the temperature range of $\SI{20}{K}\leq T \leq \SI{80}{K}$ (see the inset of Fig.~$\ref{fig4}$(b)) \cite{citation-key} and we find that the polarization is linear with respect to $E$ for $\abs{E} \leq  \SI{13.3}{kV/cm}$ in both phases.
Hence the polarization is well approximated as 
\begin{equation}
  P(T,B,E) = P_{E=0}(T,B) + \epsilon_0 \chi^E (T,B) E,
  \label{eq10}
\end{equation}
where $P_{E=0}(T,B)\equiv P(T,B,0)$ is the polarization at $E=0$, $\epsilon_0$ is the dielectric constant in vacuum, and $\chi^E$ is the electric susceptibility.
Since no phase transition is observed in the $P$-$E$ measurements, the entropy $S$ is continuous with respect to the electric field sweep for $E=0 \rightarrow E_0$.
Thus the $\DeltaSME$ in $\FZMO$ is described as follows:
\begin{equation}
\label{eq11}
    \begin{aligned}
        % \DeltaSME & = \int_0^{E_0}\left[\left(\pdv{P}{T}\right)_{B=B_0,E}-\left(\pdv{P}{T}\right)_{B=0,E}\right]\dd{E} \\
      \DeltaSME & = \int_0^{E_0}\pdv{}{T}\Bigl(P(T,B_0,E)-P(T,0,E)\Bigl)\dd{E}\\ 
      & =  \pdv{T} \Bigl(P_{E=0}(T,B_0) - P_{E=0}(T,0)\Bigl)E_0 \\
      & + \frac{\epsilon_0}{2}\pdv{T}\Bigl( \chi^E(T,B_0) - \chi^E(T,0) \Bigl)E_0^2.
    \end{aligned}
\end{equation}
Notably, in the present compound, $\partial P/\partial T$ is continuous because of the second-order phase transition \cite{matsuura2023thermodynamic} (see Fig.~\ref{fig4}(a) and its inset), and is thus well defined in both phases.

For further analysis of Eq.~(\ref{eq11}), the temperature dependences of $P_{E=0}$ and $\chi^E$ are necessary.
$P_{E=0}$ is shown in Fig.~\ref{fig4}(a), which was measured with reference to $\SI{140}{K}$ \cite{citation-key}.
$\chi^E$ is shown in Fig.~\ref{fig4}(b), which was experimentally obtained from $P$-$E$ curve measurements: $\chi^E(T,B) = \Delta P /\epsilon_0 E$, where $\Delta P = P(T,B,E) - P(T,B,0)$.
As shown in Fig.~\ref{fig4}(b), $\chi^E$ is almost independent of both the temperature and magnetic field; thus we find that the second term in Eq.~(\ref{eq11}) has a much smaller contribution to $\DeltaSME$ than the first term associated with $P_{E=0}$ \cite{citation-key}. 
By neglecting the $\chi^E$ term in Eq.~(\ref{eq11}), we derive the temperature dependence of $\DeltaSME$ for $E_0 = \SI{10}{kV/cm}$ and $B_0 = \SI{7}{T}$ using the observable macroscopic physical quantities, as displayed in Fig.~\ref{fig4}(c).
$\abs{\DeltaSME}$ takes its maximum value ($\SI{36}{mJ/kg\cdot K}$) at $T \approx \SI{57}{K}$, which corresponds to  $\sim \SI{2.6}{\%}$ of the peak value of $\abs{\DeltaSM}$ (Fig.~\ref{fig1}(f)).
The value of $\SI{36}{mJ/kg\cdot K}$ in \ce{(Fe_{0.95}Zn_{0.05})2Mo3O8} is two orders of magnitude larger than that of $\SI{0.28}{mJ/kg\cdot K}$ in the typical magnetoelectric material \ce{NdCrTiO5} estimated under the same conditions: $E_0 = \SI{10}{kV/cm}$ and $B_0 = \SI{7}{T}$ (see also Table~\ref{table1}) \cite{citation-key}.
The large difference in magnitude warrants further investigation of the origin of the multifield contribution to the MFCE in \ce{(Fe_{0.95}Zn_{0.05})2Mo3O8}, which is the subject of interest below.

\begin{figure*}[t]
    \centering
    \includegraphics[width = \hsize]{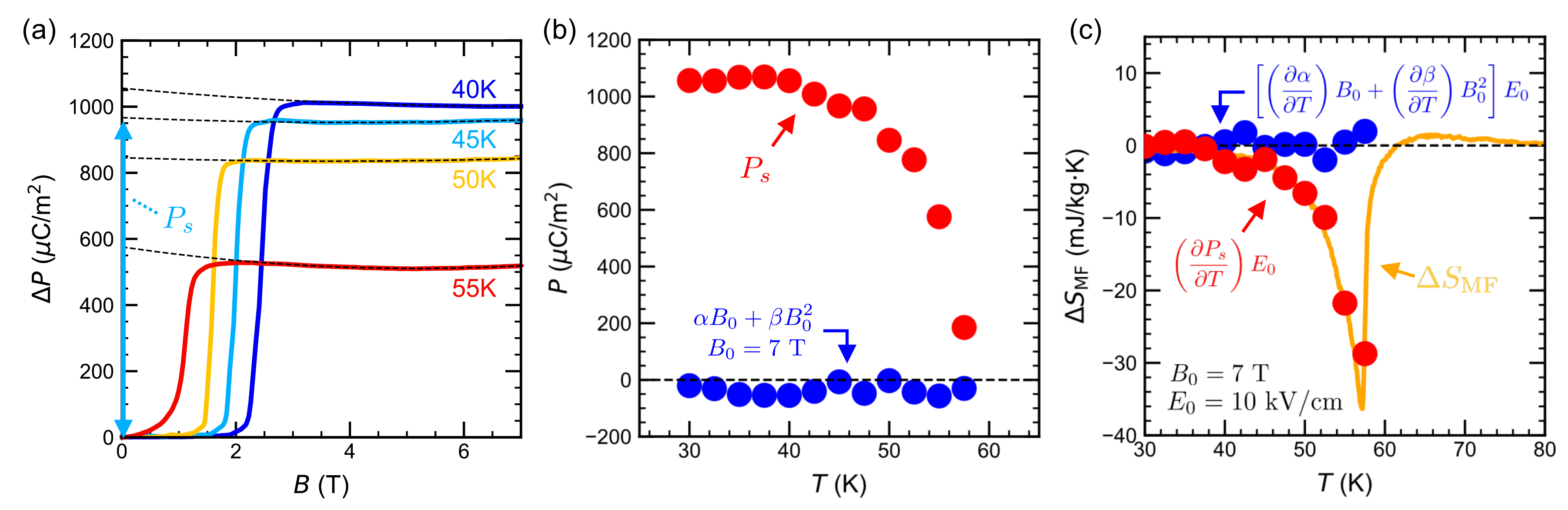}
    \caption{
    (a) Magnetic field dependence of the polarization $\Delta P = P(T,B,0) - P(T,0,0)$ at selected temperatures. The dashed lines are fitting results obtained with Eq.~\ref{eq13}.
    (b) Temperature dependences of the magnetic-order-induced polarization $P_s$ (red) and the polarization induced by the magnetoelectric coefficient $\alpha B_0 + \beta B_0^2$ (blue) at $B_0=\SI{7}{T}$ obtained from the fitting with Eq.~(\ref{eq13}).
    Below $\SI{50}{K}$, the first-order phase transition between the AFM and FRI phases occurs during the field sweep to $\SI{7}{T}$.
    Between $\SI{50}{K}$ and $\SI{56}{K}$, close to the triple point of the three phases (AFM, PM, and FRI phases), the phase transition from the AFM to FRI phases via the PM phase occurs.
    (c) Temperature dependence of the multifield contribution to the MFCE from magnetic-order-induced polarization (red markers), from the polarization induced by the linear and second-order magnetoelectric effect (blue markers), and the total $\DeltaSME$ at $B_0 = \SI{7}{T}$ and $E_0 = \SI{10}{kV/cm}$ (orange).}
    \label{fig5}
\end{figure*}

In the present compound, $P_{E=0}(T,B_0) = P(T,B_{0},0)$ in the FRI phase with reference to $\SI{0}{T}$ can be described by
\begin{equation}
    P_{E=0}(T,B_0) - P_{E=0}(T,0)  =  P_s (T) + \alpha (T) B_{0} + \beta (T) B_{0}^2,
    \label{eq13}
\end{equation}
where $P_s(T)$ represents the magnetic-order-induced polarization in the FRI phase obtained from the extrapolation to $B=\SI{0}{T}$ (see also Fig.~\ref{fig5}(a)), $\alpha(T)$ represents the linear magnetoelectric coefficient, and $\beta(T)$ denotes the second-order magnetoelectric coefficient \cite{kurumaji2015doping,kurumaji2017diagonal}.
To scrutinize the contribution from each term in Eq.~(\ref{eq13}), $P$-$B$ curves are measured, as shown in Fig.~\ref{fig5}(a).
At $T=\SI{45}{K}$, for instance, $P_s = \SI{967}{\mu C/m^2}$, $\alpha = \SI{-9.2}{ps/m}$, and $\beta = \SI{1.4e-18}{s/A}$ are obtained (dotted line at $\SI{45}{K}$ in Fig.~\ref{fig5}(a)). 
Based on the fitting with Eq.~(\ref{eq13}), we show the temperature dependences of the magnetic-order-induced polarization $P_s$ and the linear and second-order magnetoelectric effect-induced polarization $\alpha B_0 + \beta B_0^2$ at $B_0=\SI{7}{T}$ in Fig.~\ref{fig5}(b).
$P_s$ pronouncedly decreases toward $\SI{57.5}{K}$ and disappears above $\SI{60}{K}$ \cite{citation-key}.
$P_{s}$ accounts for most of the polarization in the FRI phase of \ce{(Fe_{0.95}Zn_{0.05})2Mo3O8}, whereas $\alpha B_0 + \beta B_0^2$ is $\sim 10^{-2}$ times smaller than $P_s$, i.e., $ P_{E=0}(T,B_0) - P_{E=0}(T,0)  \approx  P_s (T)$.
Thus we find that $\DeltaSME$ is well described by $\DeltaSME \approx (\partial P_s/\partial T)E_0$.
Figure~\ref{fig5}(c) displays the temperature dependence of the multifield contribution to the MFCE estimated from $P_s$ (shown as red circles), and the magnetoelectric effect (blue circles), together with the result of Fig.~\ref{fig4}(c) obtained from Eq.~(\ref{eq11}).
The total multifield contribution $\DeltaSME$ and $(\partial P_s/\partial T)E_0$, derived from the different approaches, show fairly good quantitative agreement, whereas the contribution from the linear
and second-order magnetoelectric effect-induced polarization, $[(\partial \alpha/\partial T) B_0 + (\partial \beta/\partial T) B_0^2]E_0$, is one order of magnitude smaller than that from $P_s$.
These results indicate that the temperature-dependent $P_s$ is the dominant factor for the observed multifield contribution to the MFCE.

\section{Discussion}
Finally, we discuss the magnitude of $\DeltaSME$ in $\FZMO$ with reference to other magnetoelectric compounds. The available experimental data are extracted from previous reports and used to calculate the first term in Eq.~(\ref{eq11}), assuming that the contribution from the second term associated with $\chi^E$ in Eq.~(\ref{eq11}) is negligible, as in the case of $\FZMO$ \cite{citation-key}.  
The maximum values of $|\DeltaSME|$ in various magnetoelectric multiferroics are shown in Table~\ref{table1}.
Materials that exhibit magnetic-order-induced polarization, such as \ce{TbMnO3}, \ce{MnWO4}, \ce{Ba2CoGe2O7}, \ce{Ni3V2O8}, \ce{GdMn2O5}, \ce{Fe2Mo3O8}, and \ce{(Fe_{0.95}Zn_{0.05})2Mo3O8}, tend to have larger multifield contributions than linear-magnetoelectric materials, such as \ce{NdCrTiO5} and \ce{Cr2O3},  although the linear-magnetoelectric material \ce{Co4Nb2O9} is an exception.
This comparison implies that magnetic-order-induced polarization plays a significant role in realizing a large multifield contribution to the MFCE.

\begin{table}[tb]
\caption{Multifield contribution to the MFCE in various multiferroics. $\Delta S_{\rm MF}$ is calculated at $B=B_0$ and $E_0=\SI{10}{kV/cm}$, and the maximum value is shown.
``Linear ME effect'' and ``Magnetic-order'' in the fourth column represent the linear magnetoelectric effect and magnetic-order-induced polarization, respectively.}
\label{table1}
\begin{ruledtabular}
\begin{tabular}{lccccl}
 &$T$ & $B_0$  & $\abs{\Delta S_{\rm MF}}$  & Main mechanism \\[0.5ex]
 &($\SI{}{K}$) & ($\SI{}{T}$)  & ($\SI{}{mJ/kg\cdot K}$) & of polarization \\[0.5ex]
\hline
\ce{(Fe_{0.95}Zn_{0.05})2Mo3O8} & $56$ & $7$ & $36$  & Magnetic-order \\
\ce{NdCrTiO5} \cite{PhysRevB.85.024415} & $20$ & $7$ & $0.28$ & Linear ME effect \\
\ce{Cr2O3} \cite{PhysRevB.87.180408} & $307$ & $9$ & $3.0$ & Linear ME effect \\
\ce{Co4Nb2O9} \cite{khanh2016magnetoelectric} & $26$ & $7$ & $24$ & Linear ME effect \\
\ce{Ba2CoGe2O7} \cite{yi2008multiferroicity} & $6.7$ & $8$ & $24$ & Magnetic-order \\
\ce{MnWO4} \cite{PhysRevB.74.184431} &$7.6$ & $9$ & $25$ & Magnetic-order \\
\ce{TbMnO3} \cite{kimura2005magnetoelectric} & $26$ & $7$ & $26$ & Magnetic-order \\
\ce{Ni3V2O8} \cite{PhysRevLett.95.087205} & $4.0$ & $5$ & $53$ & Magnetic-order \\
\ce{Fe2Mo3O8} \cite{kurumaji2015doping} & $40$ & $9$ & $62$ & Magnetic-order \\
\ce{GdMn2O5} \cite{PhysRevB.94.174446} & $28$ & $8$ & $73$ & Magnetic-order \\
\end{tabular}
\end{ruledtabular}
\end{table}

\section{Conclusion}
In the present study, we conducted quantitative evaluations of the multifield contribution to the MFCE in the magnetoelectric compound $\FZMO$. 
We obtain the multifield contribution $\DeltaSME =\SI{-36}{mJ/kg\cdot K}$ at $B_0 = \SI{7}{T}$ and $E_0 = \SI{10}{kV/cm}$, which is two orders of magnitude larger than the $\DeltaSME$ in the well-known linear magnetoelectric material \ce{NdCrTiO5}.
The large difference in magnitude between $\FZMO$ and \ce{NdCrTiO5} can be attributed to the presence of magnetic-order-induced polarization in $\FZMO$.
Our study promotes the understanding of thermodynamics in multiferroic systems and is helpful for constructing a strategy for designing caloric materials with multiple degrees of freedom.

\begin{acknowledgments}
K.I. and K.M. acknowledge Dr.~M.~Kriener for assisting with the high-vacuum mode in the adiabatic temperature measurement.
This work was supported by JSPS KAKENHI Grant No. JP23K13054 and No. JP24K08227.
\end{acknowledgments}

\section*{Author contribution}
K.M. and F.K. designed the research;
K.I. and K.M. performed the research; K.I. and K.M. analyzed the data; K.I., K.M. and F.K. wrote the paper; T.N. commented on the experimental setup and analysis; T.K. and Y.T. provided the samples.
K.I. and K.M. contributed equally to this work.

\section*{DATA AVAILABILITY}
The data are available from the authors upon reasonable request.

% \appendix
\renewcommand{\thefigure}{A\arabic{figure}}
\renewcommand{\theequation}{A\arabic{equation}}
\setcounter{figure}{0}
\setcounter{equation}{0}

\section*{Appendix A: Thermodynamic details of the multifield caloric effect and multiorder monocaloric effect}
\subsection*{Multifield caloric effect}
To discuss the MFCE, we adopt the Gibbs free energy $G(T, B, E)$ as a thermodynamic potential with arguments of temperature $T$, external magnetic field $B$ and external electric field $E$, which are directly controllable parameters in experiments.
From the relation $\dd{G} = - S\dd{T} - M\dd{B} - P\dd{E}$, the entropy $S = S(T,B,E)$, magnetization $M = M(T,B,E)$, and polarization $P=P(T,B,E)$ are defined as 
\begin{equation}
\left\{ \,
\begin{aligned}
    S(T,B,E) &= -\left(\pdv{G}{T}\right)_{B,E} \\ 
    M(T,B,E) &= -\left(\pdv{G}{B}\right)_{T,E} \\
    P(T,B,E) &= -\left(\pdv{G}{E}\right)_{T,B}.
\end{aligned}
\right.
\end{equation}
The magnetoelectric susceptibility, which describes the cross-correlation between $M$ and $E$ (or $P$ and $B$), is given by
\begin{equation}
    \alpha_{\rm ME} = -\left(\pdv{G}{B}{E}\right)_T = \left(\pdv{M}{E}\right)_{T,B} = \left(\pdv{P}{B}\right)_{T,E}.
\end{equation}
$\alpha_{\rm ME} = \alpha_{\rm ME}(T,B,E)$ depends on both magnetic and electric fields and, in this sense, represents a more general magnetoelectric susceptibility than the linear magnetoelectric coefficient $\alpha \equiv \alpha_{\rm ME}(T,B=0,E=0)$.
When both magnetic and electric fields are varied from zero at a fixed temperature, the entropy changes accordingly. Using Taylor expansion, the resulting isothermal entropy change $\Delta S_{\rm iso}$ can be expressed as
\begin{equation}
\begin{aligned}
    \Delta S_{\rm iso} =& S(T,B, E) - S(T,0,0) \\
    = &\sum_{n=0}^{\infty} \frac{1}{n!}\left[ B \left(\pdv{B}\right) + E \left(\pdv{E}\right) \right]^n S(T,B,E) \\
     = & \sum_{n=0}^{\infty} \frac{1}{n!} \left(\pdv[n]{S}{B}\right) B^n  + \sum_{n=0}^\infty \frac{1}{n!}\left(\pdv[n]{S}{E}\right) E^n  \\
     &+ \sum_{n=2}^{\infty}  \sum_{k=1}^{n-1} \frac{1}{n!}\binom{n}{k}  \left(\frac{\partial ^n S}{\partial B^k \partial E^{n-k}}\right) B^k E^{n-k}.
\end{aligned}
\label{Taylor}
\end{equation}
The first term corresponds to the isothermal entropy change in the magnetocaloric effect $\DeltaSM$, which can be measured by applying a magnetic field alone ($B\neq0$ and $E=0$): 
\begin{equation}
    \DeltaSM =  S(T,B,0)-S(T,0,0) =  \sum_{n=0}^{\infty} \frac{1}{n!} \left(\pdv[n]{S}{B}\right) B^n.
\end{equation}
The second term corresponds to the isothermal entropy change in the electrocaloric effect $\DeltaSE$, which can be measured by applying an electric field alone ($B=0$ and $E\neq0$):
\begin{equation}
    \DeltaSE = S(T,0,E)-S(T,0,0) =\sum_{n=0}^{\infty} \frac{1}{n!} \left(\pdv[n]{S}{E}\right) E^n.
\end{equation}
Note that both $\DeltaSM$ and $\DeltaSE$ are categorized as monocaloric effects. The third term in Eq.~(\ref{Taylor}) represents the cross term between the magnetic and electric fields, and this term is absent in monocaloric effects by definition. The third term, $\DeltaSME$, can therefore be considered unique to the MFCE. Furthermore, $\DeltaSME$ is related to $\alpha_{\rm ME}$, as explicitly shown by
\begin{equation}
\begin{aligned}
\Delta S_{\rm MF} &= \sum_{n=2}^{\infty}  \sum_{k=1}^{n-1} \frac{1}{n!} \binom{n}{k}  \left(\frac{\partial ^n S}{\partial B^k \partial E^{n-k}}\right) B^k E^{n-k}\\ &= - \sum_{n=2}^{\infty} \sum_{k=1}^{n-1}  \frac{1}{n!} \binom{n}{k} \left[\frac{\partial ^n }{\partial B^k \partial E^{n-k}} \left(\pdv{G}{T}\right)\right] B^k E^{n-k} \\
&=  \sum_{n=2}^{\infty}\sum_{k=1}^{n-1} \frac{1}{n!} \binom{n}{k}  \left[\frac{\partial ^{n-2}}{\partial B^{k-1} \partial E^{n-k-1} }\left(\pdv{\alpha_{\rm ME}}{T}\right) \right] B^k E^{n-k}.
\end{aligned}
\label{ME_term_taylor}
\end{equation}
$\DeltaSME$ is therefore a good quantity for representing the isothermal entropy change due to the cross-correlational coupling in the MFCE.
Specifically, the above arguments using Taylor expansion do not hold when a phase transition is involved in the considered $(T, B, E)$ space. Nevertheless, $\DeltaSME \equiv \Delta S_{\rm iso} - \DeltaSM - \DeltaSE$ remains well defined even if a phase transition is involved, as discussed in the main text and Refs.~\cite{manosa2023cross,stern2018multicaloric}.

\subsection*{Multiorder monocaloric effect}
We discuss the isothermal entropy change in a monocaloric experiment for a system characterized by multiple order parameters, such as magnetoelectric multiferroics.
In magnetoelectric multiferroics, both the magnetization and polarization are simultaneously varied when a single external field is applied; therefore, expecting that the electricity-magnetism-coupled degrees of freedom affect the resulting isothermal entropy change even in a monocaloric experiment is reasonable. In fact, there are several reports related to this issue \cite{vopson2012multicaloric,vopson2013theory,VOPSON201614,edstrom2020prediction,planes2014thermodynamics,li2024cross,andrade2019multicaloric,10.1063/1.5090599,stern2017giant,starkov2015rebuttalthemulticaloriceffect,MENG2013567}.
As discussed in the previous subsection, the $\DeltaSME$ that appears in the MFCE is irrelevant to the monocaloric effect by definition.
The question that arises here is how the characteristics of the coupled degrees of freedom manifest themselves in the isothermal entropy change in monocaloric experiments.

To address this issue, we consider a magnetocaloric effect in magnetoelectric multiferroics and adopt the Helmholtz free energy $F(T,M,P)$ as a thermodynamic potential, in which $T$, $M$, and $P$ are arguments.
The total differential of $F(T,M,P)$ is given as $\dd{F} = - S\dd{T} + B\dd{M} + E\dd{P}$; thus, $S = S(T,M,P)$, $B = B(T,M,P)$, and $E = E(T,M,P)$ are given by
\begin{equation}
\left\{ \,
\begin{aligned}
    S(T,M,P) &= -\left(\pdv{F}{T}\right)_{M,P} \\
    B(T,M,P) &= \left(\pdv{F}{M}\right)_{T,P} \\
    E(T,M,P) &= \left(\pdv{F}{P}\right)_{T,M}.
\end{aligned}
\right.
\end{equation}
The total differential of the entropy $S(T, M, P)$ is given as 
\begin{equation}
    \dd{S} = \left(\pdv{S}{T}\right)_{M,P} \dd{T} + \left(\pdv{S}{M}\right)_{T,P} \dd{M} +  \left(\pdv{S}{P}\right)_{T,M} \dd{P}.
    \label{MCE_total}
\end{equation}
Given $\dd{T}=0$ in an isothermal process, the isothermal entropy change in a differential form, $\dd{S}_{\rm iso}$, is given by
\begin{equation}
    \dd{S}_{\rm iso} = \left(\pdv{S}{M}\right)_{T,P} \dd{M} +  \left(\pdv{S}{P}\right)_{T,M} \dd{P}.
    \label{MCE_Fform}
\end{equation}
In the case of magnetoelectric multiferroics, the second term involving $\dd{P}$ is generally finite in magnetocaloric experiments.
Given that a directly controllable parameter in magnetocaloric experiments is $B$, not $M$ or $P$, rewriting Eq.~(\ref{MCE_Fform}) using $\dd{B}$ is helpful.
When $M$ and $P$ are viewed as functions of $T$, $B$, and $E$, the total differentials of $M$ and $P$ are given as 
\begin{equation}
\begin{aligned}
\dd{M} &= \left(\pdv{M}{T}\right)_{B,E} \dd{T} 
+ \left(\pdv{M}{B}\right)_{T,E} \dd{B} + \left(\pdv{M}{E}\right)_{T,B} \dd{E},\\
\dd{P} &= \left(\pdv{P}{T}\right)_{B,E} \dd{T} 
+ \left(\pdv{P}{B}\right)_{T,E} \dd{B} + \left(\pdv{P}{E}\right)_{T,B} \dd{E}.
\end{aligned}
\label{dM_dP_totaldiff}
\end{equation}
In magnetocaloric experiments, $T$ and $E$ (external electric field) are fixed, i.e., $\dd{T}=0$ and $\dd{E}=0$. Thus, Eq.~(\ref{dM_dP_totaldiff}) reduces to $\dd{M} = (\partial M/\partial B)_{T,E} \dd{B}$ and
% \sout{$\dd{P} = (\partial P/\partial B)_{T,E} \dd{B}$} 
$\dd{P} = (\partial P/\partial B)_{T,E} \dd{B} = \alpha_{\rm ME} \dd{B}$. 
Equation (\ref{MCE_Fform}) is eventually rewritten as
\begin{equation}
    \begin{aligned}
    \dd{S}_{\rm iso} 
        & = \left[ \left(\pdv{S}{M}\right)_{T,P} \left(\pdv{M}{B}\right)_{T,E} +   \left(\pdv{S}{P}\right)_{T,M} \alpha_{\rm ME} \right] \dd{B}.
    \end{aligned}
    \label{last_result}
\end{equation}
The second term, $\left(\partial{S}/\partial{P}\right)_{T,M} \alpha_{\rm ME} \dd{B}$, explicitly involves the magnetoelectric susceptibility and may therefore be viewed as representing the characteristics of the coupled degrees of freedom in the magnetocaloric effect.
To determine $(\partial S/\partial P)_{T,M} = -(\partial E/\partial T)_{P,M}$ (Maxwell's relation) through actual experiments, the temperature dependence of the external electric field such that $P$ and $M$ are maintained must be measured. Note, however, that this is not easily implemented for electricity-magnetism-coupled systems because independently controlling $T$, $M$, and $P$ is experimentally challenging.
Thus, although according to the equations, the magnetocaloric effect in multiferroics clearly involves a unique term in the isothermal entropy change, we do not consider experimental determination of this contribution.
In the main text, we consider the MFCE instead and extract the term unique to the MFCE, i.e., multifield contribution $\DeltaSME$.

Finally, we mention the relationship between the thermodynamic formulas above and the Landau approach for the MOCE.
Recently, regarding the electrocaloric effect in multiferroic SrMnO$_{3}$ \cite{edstrom2020prediction} and BiFeO$_{3}$ \cite{li2024cross}, which is classified according to our definition as the MOCE, a large contribution of the coupled degrees of freedom was theoretically proposed on the basis of Landau's pseudo-free energy.
Following the literature \cite{edstrom2020prediction}, below, we reproduce the arguments based on Landau's pseudo-free energy, $F_L(T,M,P)$, which is given by  
\begin{equation}
\begin{aligned}
        F_L(T,M,P) =& \frac{1}{2}a_M(T-T^M_0) M^2 + \frac{1}{2}a_P(T-T^P_0) P^2 \\
        & + \frac{1}{4}b_M M^4 + \frac{1}{4}b_P P^4 + \frac{1}{2}\lambda M^2 P^2,
\end{aligned}
\end{equation}
where $T^M_0$ and $T^P_0$ denote the critical temperatures and $\lambda$ is the coupling parameter between $M$ and $P$.
In the Landau approach, $P$ and $M$ are implicitly assumed to be small and thus higher-order terms are often omitted.
We consider a magnetocaloric effect under $E=0$.
The Gibbs free energy $G(T,B)$ of this model is given by the following Legendre transformation of $F_{L}$:
\begin{equation}
G(T,B) = \min_{M,P} [F_L(T,M,P) - MB].
\end{equation}
By calculating the stationary points with respect to $M$ and $P$, i.e., $(\partial F_L/\partial M)_{T,P} - B = (\partial F_L/\partial P)_{T,M}=0$, the equations that $M=M(T,B)$ and $P=P(T,B)$ should satisfy are obtained as follows:
\begin{equation}
\left\{ \,
\begin{aligned}
    M \left[ a_M(T-T^M_0)+ b_M M^2 + \lambda P^2 \right] - B & = 0 \\
     P=0 \quad \mathrm{or} \quad a_P(T-T^P_0) + b_P P^2 + \lambda M^2 &= 0 .
\end{aligned}
\right.
\label{minimum_eq}
\end{equation}
Hence, the Gibbs free energy with $P \neq 0$ is given by
\begin{equation}
\begin{aligned}
   G(T,B) =& -\frac{a_P^2}{4b_P}(T-T^P_0)^2 \\
   &+ \frac{1}{2}\left[a_M(T-T^M_0)- \frac{\lambda a_P}{b_P}(T-T^P_0)\right]M(T,B)^2 \\
   &+ \frac{1}{4}\left(b_M - \frac{\lambda^2}{b_P}\right)M(T,B)^4- BM(T,B). 
\end{aligned}
\end{equation}
The entropy $S=S(T,B)$ is then obtained via $S=-(\partial G/\partial T)_{B}$, leading to
\begin{equation}
\begin{aligned}
       S(T,B) &= \frac{a_P^2}{2b_P}(T-T^P_0) - \frac{1}{2}\left( a_M - \frac{\lambda a_P}{b_P}\right)M(T,B)^2 \\
       & = -\frac{1}{2}a_M M(T,B)^2 - \frac{1}{2}a_P P(T,B)^2.
\end{aligned}
\label{Landau_S}
\end{equation}
To compare this result with Eq.~(\ref{last_result}), we derive $\dd{S}_{\rm iso}$, the differential form of Eq.~(\ref{Landau_S}),
\begin{equation}
    \begin{aligned}
    \dd{S}_{\rm iso} &\equiv \left(\pdv{S}{B}\right)_{T} \dd{B}\\
    &= \left[ -a_M M \left(\pdv{M}{B}\right)_{T} - a_P P \left(\pdv{P}{B}\right)_{T} \right] \dd{B}\\
    & = \left[ \left(\pdv{S}{M}\right)_{T,P} \left(\pdv{M}{B}\right)_{T,E} + \left(\pdv{S}{P}\right)_{T,M} \alpha_{\rm ME} \right] \dd{B},
\end{aligned}
\label{Landau_dS}
\end{equation}
where $-a_M M = (\partial S/\partial M)_{T, P}$ and $-a_P P = (\partial S/\partial P)_{T, M}$ are calculated via Eq.~(\ref{Landau_S}).
Note that the last line of Eq.~(\ref{Landau_dS}) is the same as the last line of Eq.~(\ref{last_result}). Thus, regarding the MOCE, the conclusions drawn from the Landau approach are consistent with those drawn from thermodynamics alone, as expected.

\renewcommand{\thefigure}{B\arabic{figure}}
\renewcommand{\theequation}{B\arabic{equation}}
\setcounter{figure}{0}
\setcounter{equation}{0}

\section*{Appendix B: Magnetocaloric effect measurement}
\begin{figure}[t]
  \centering
  \includegraphics[width=12cm]{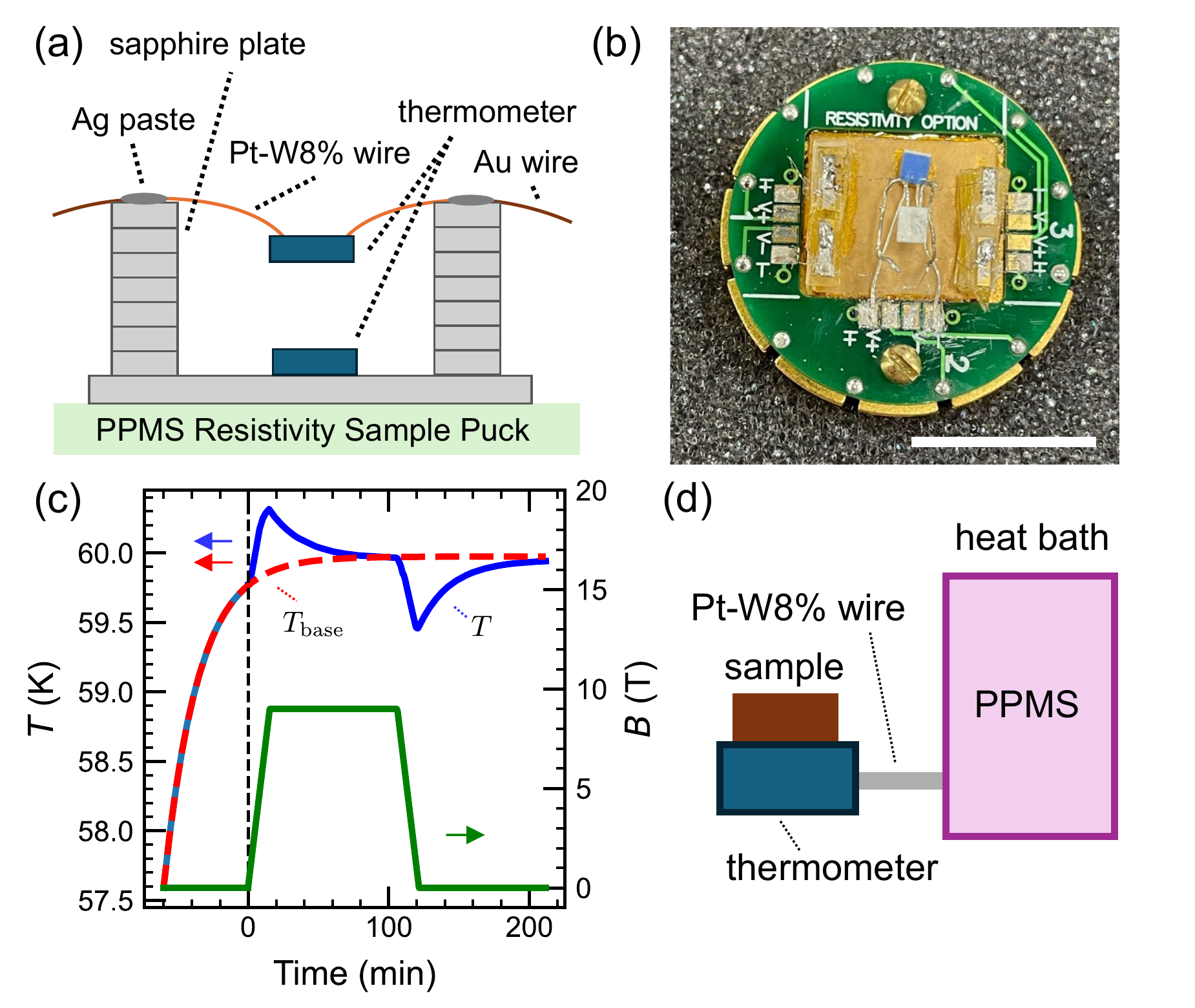}
  \caption{
  (a) Schematic illustration of the setup for the direct adiabatic temperature change measurement (side view).
  (b) Photograph of the actual setup. The white bar represents $\SI{10}{mm}$.
    (c) Time profile of the sample temperature $T$ (blue solid line) and magnetic field (green solid line) at $\SI{60}{K}$ under adiabatic conditions. $T_{\mathrm{base}}$ (red dashed line) is the fitting obtained with the exponential function (Eq.~(\ref{Eq_Tbase})). 
  (d) Thermal conductivity model of the present setup for analyzing the thermal diffusion from wires.
  }
  \label{setup}
\end{figure}

The adiabatic temperature change was measured by our home-build setup on the PPMS resistivity puck.
Figures \ref{setup}(a) and \ref{setup}(b) show a schematic illustration of the direct adiabatic temperature measurement setup and a photograph of the actual setup, respectively.

\begin{figure}[t]
  \centering
  \includegraphics[width=12cm]{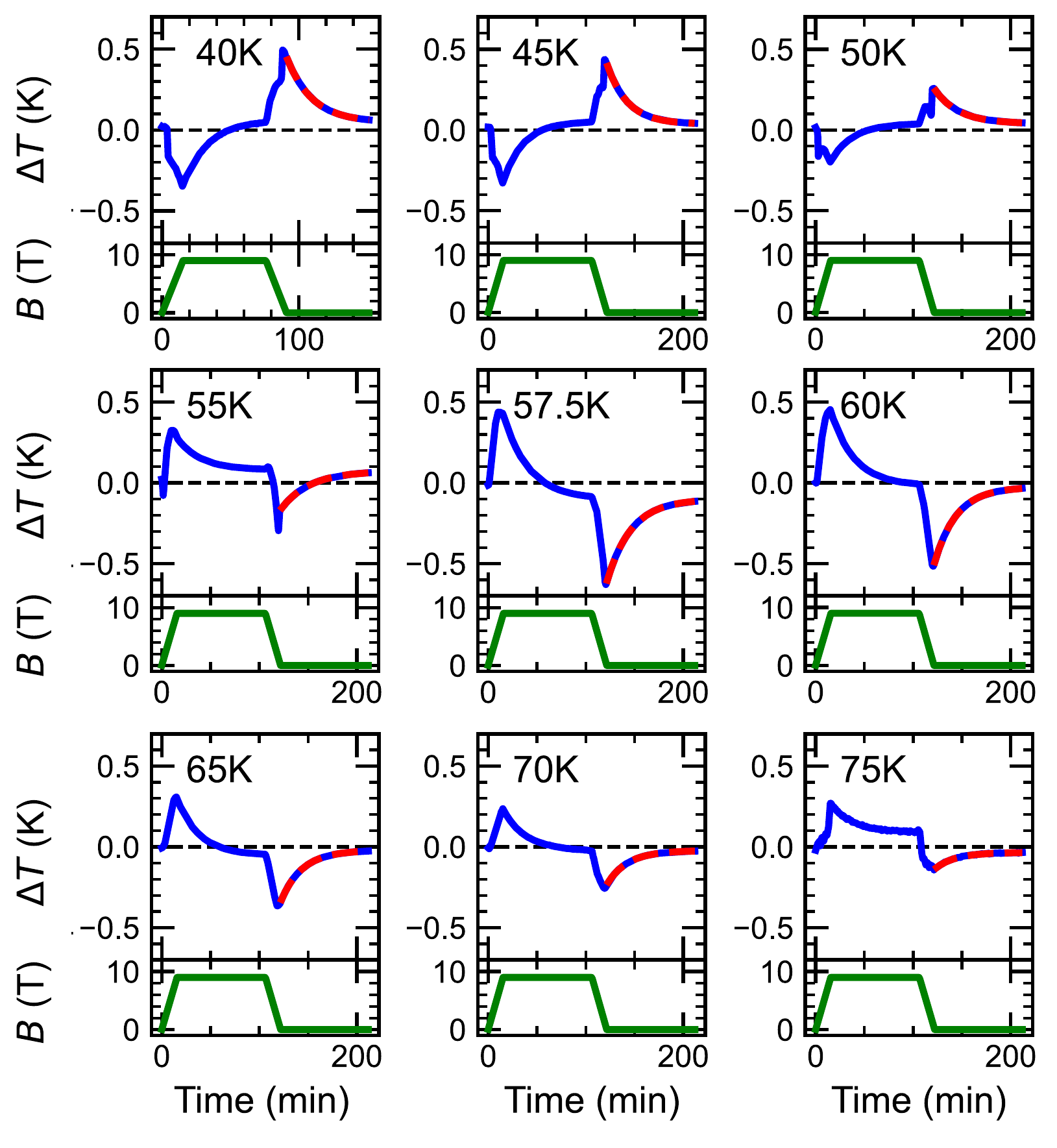}
  \caption{
  Time profiles of sample temperature changes $\Delta T$ in the adiabatic temperature measurements with reference to $T_{{\rm base}}$ from $\SI{40}{K}$ to $\SI{75}{K}$ (blue solid line) and of the magnetic fields (green solid line). 
  The exponential curve fittings (Eq.~(\ref{tau_eq})) for estimating the thermal diffusion from wires are shown as red dashed lines.
  }
  \label{MCE_timedep}
\end{figure}

In the present study, only time profiles of the sample temperature were analyzed to obtain the adiabatic temperature change $\Delta T_{{\rm ad}}$.
To accurately measure the adiabatic sample temperature change $\Delta T_{{\rm ad}}$, which is the temperature change of the sample when magnetic fields are adiabatically applied or removed, the following two problems need to be considered: (i) drift of the base temperature and (ii) thermal diffusion from the sample to the thermal bath during the field sweep. 
To address problem (i), we checked the time profile of the sample temperature $T(t)$ at $\SI{60}{K}$ (Fig.~\ref{setup}(c)).
Before the magnetic field was applied, we waited $\SI{60}{min}$ for the sample temperature to settle at approximately $\SI{60}{K}$. 
During this waiting period ($t\leq0$ in Fig.~\ref{setup}(c)), the sample temperature $T(t)$ (blue solid line in Fig.~\ref{setup}(c)) increased from $\SI{57.5}{K}$ to $\SI{59.8}{K}$, which is quite close to the target temperature of $\SI{60}{K}$.
Nevertheless, a slight difference between $T$ ($t=\SI{0}{min}$) and $T$ ($t=\SI{200}{min}$) can still be seen. 
This difference occurs because the settling time ($\sim \SI{60}{min}$) is not sufficient to completely stabilize the sample temperature in the high-vacuum space; thus the sample temperature slightly changes during the scan.
Hence, to estimate this temperature drift in $T(t)$, the following equation was applied to $T(t)$ ($t\leq0$): 
\begin{equation}
  T_{{\rm base}}(t) = T_{0} - a\exp\left(\frac{t}{\tau_{0}}\right), 
  \label{Eq_Tbase}
\end{equation}
where $T_{0}$, $a$ and $\tau_{0}$ are fitting parameters. 
The result of this fitting at $\SI{60}{K}$ is shown as the red dashed line in Fig.~\ref{setup}(c). 
The sample temperature change $\Delta T$ can then be measured with reference to $T_{{\rm base}}(t)$: $\Delta T(t) = T(t) - T_{\mathrm{base}}(t)$.
Figure~\ref{MCE_timedep} summarizes the time evolution of $\Delta T$ (blue solid lines in Fig.~\ref{MCE_timedep}) at various temperatures when magnetic fields are swept.

\begin{figure}[t]
  \centering
  \includegraphics[width=12cm]{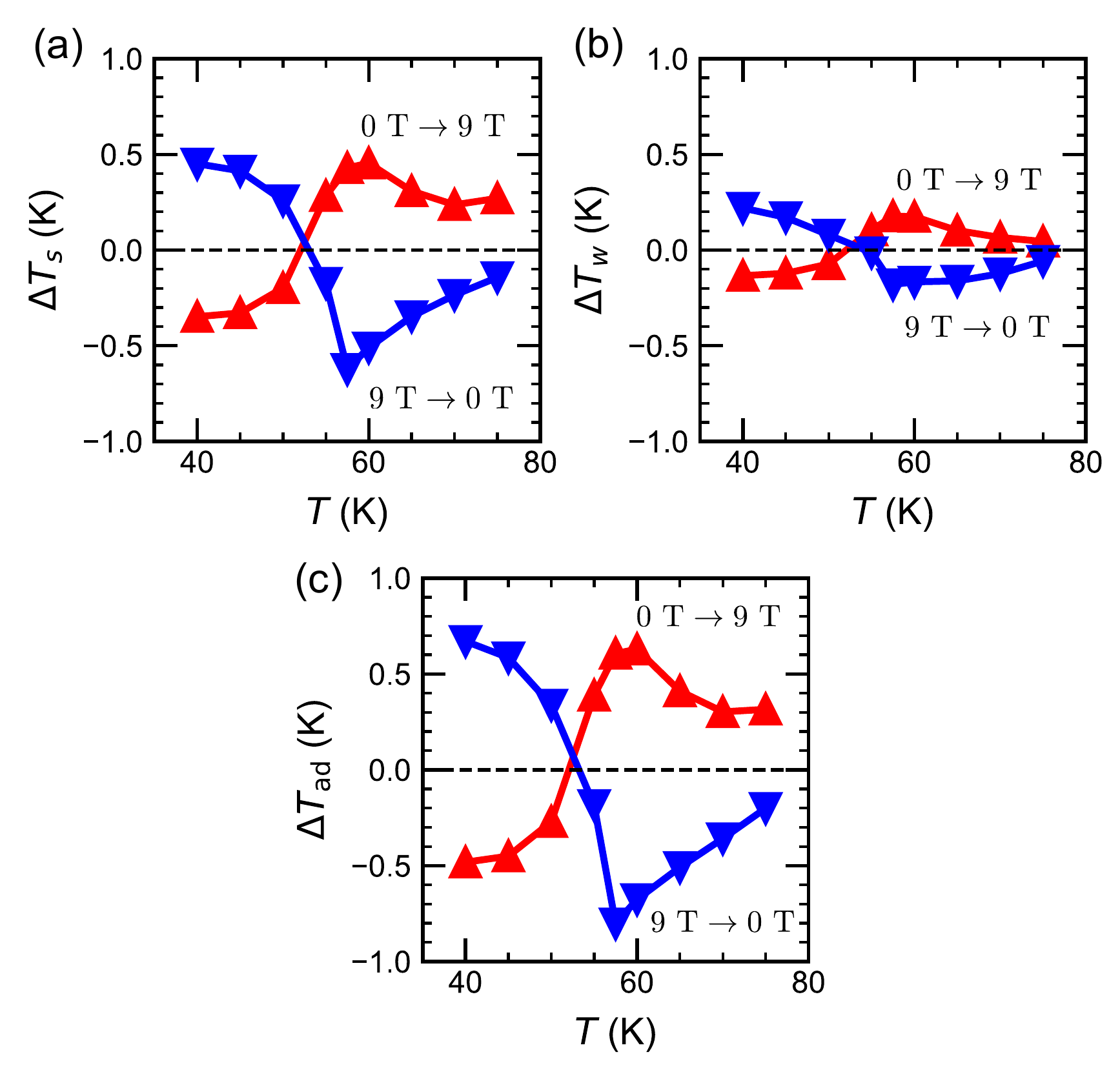}
  \caption{
  Temperature dependences of (a) $\Delta T_{s}$, (b) $\Delta T_{w}$, and (c) $\Delta T_{{\rm ad}}$ (see Eq.~(\ref{Tad_wire_eq})). The red markers are from the field-applying process ($\SI{0}{T} \rightarrow \SI{9}{T}$) and the blue markers are from the field-removing process ($\SI{9}{T} \rightarrow \SI{0}{T}$).
  }
  \label{MCE_Tdep}
\end{figure}

Next, we consider (ii) the effect of thermal diffusion from the sample to the thermal bath during magnetic field sweeping. 
The heat generated from the sample invariably diffuses through the wires (Pt-W $\SI{8}{\%}$) toward the heat bath; thus, this diffusion affects the evaluation of $\Delta T_{{\rm ad}}$.
Figure~\ref{setup}(d) shows the thermal conductivity model considered in our analysis.
Considering the heat balance, i.e., the amount of heat going in and out of the system, the following equation is obtained \cite{yonezawa2013first}:
\begin{equation}
\begin{aligned}
  \pdv{S}{B}&  = -\frac{C_s}{T}\left(\dv{T}{B}\right) -\frac{k(T-T_{\mathrm{base}})}{T\dot{B}}  \\
  & = -\frac{C_s}{T}\left(\dv{T}{B}\right) -\frac{k\Delta T(t)}{T\dot{B}}, 
  % \tag{S2}
  \label{dSdB}
\end{aligned}
\end{equation}
where $S$ represents the entropy of the sample, $C_s$ is the heat capacity of the sample, $k$ is referred to as the thermal conductance of the wire, and $\dot{B}=\mathrm{d}B/\mathrm{d}t$.
Here, we assume that the heat capacity of the thermometer $C_t$ is negligibly small.
The heat generated from the thermometer when measuring the resistance, $P=RI^2$, is only $\simeq \SI{2}{\mu W}$; thus its contribution is negligible and not included in Eq.~(\ref{dSdB}).
Using Eq.~(\ref{dSdB}), $\Delta T_{\mathrm{ad}}$ can be further expanded as follows:
\begin{equation}
  \begin{aligned}
  \DeltaTad &= -\int_0^{B_0} \frac{T}{C_s}\left(\pdv{S}{B}\right)\dd{B} \\
  & = \int_0^{B_0} \left(\dv{T}{B}\right)\dd{B} +\int_0^{B_0} \frac{k\Delta T}{C_s\dot{B}}\dd{B}  \\ 
  &  = \int_{T(B=0)}^{T(B=B_0)} \dd{T}  + \int_{t(B=0)}^{t(B=B_0)} \frac{k\Delta T}{C_s}\dd{t}  \\
  &= \Delta T_s + \int_{t(B=0)}^{t(B=B_0)} \frac{k\Delta T}{C_s}\dd{t},  
\label{Tad_model}
\end{aligned}
\end{equation}
where $\Delta T_s = T(B=B_0) - T(B=0)$ is the sample temperature change from $B=0$ to $B_0$.
Consequently, if the magnetic field dependences of $k$ and $C_s$ are sufficiently weak, $\DeltaTad$ can be further simplified:
\begin{equation}
  \label{Tad_wire_eq}
  \Delta T_{\mathrm{ad}} = \Delta T_s + \frac{k}{C_s}\int_{t(B=0)}^{t(B=B_0)} \Delta T \dd{t}
  =\Delta T_s + \Delta T_w.
  % \tag{S4}
\end{equation}
$\Delta T_w$ in Eq.~(\ref{Tad_wire_eq}) is the correction due to thermal diffusion.
$k/C_s$ was estimated from the fitting of the $\Delta T(t)$ data after the removal of the magnetic fields: 
\begin{equation}
\label{tau_eq}
    \Delta T(t) = \Delta T_0 \exp\left(-\frac{t-t_{{\rm r}}}{\tau}\right) \ \ \ \  (t_{r} \le t),
    % \tag{S5}
\end{equation}
where $t_{{\rm r}}$ denotes the time when the magnetic field was removed, and $\Delta T_0$ and $\tau$ are fitting parameters. 
The results of the fitting with Eq.~(\ref{tau_eq}) are shown as red dashed lines in Fig.~\ref{MCE_timedep}.
To be precise, $\tau$ obtained from the fitting includes the contribution from $C_{t}$, but here, we consider $\tau \simeq C_s/k$ because $C_t \ll C_s$.

Figures \ref{MCE_Tdep}(a) and (b) show the temperature dependences of $\Delta T_{\rm s}$ and $\Delta T_w$ in Eq.~(\ref{Tad_wire_eq}) at $B_{0} = \SI{9}{T}$, respectively.
We thus obtained the temperature dependence of the adiabatic temperature change $\Delta T_{{\rm ad}}$ (Fig.~\ref{MCE_Tdep}(c)).

%%%% sup
\renewcommand{\thefigure}{S\arabic{figure}}
\renewcommand{\theequation}{S\arabic{equation}}
\setcounter{figure}{0}
\setcounter{equation}{0}

\clearpage

\section*{Supplemental  Material}
\subsection*{Polarization measurements in $\FZMO$}
\begin{figure}[h]
  \centering
  \includegraphics[width=\hsize]{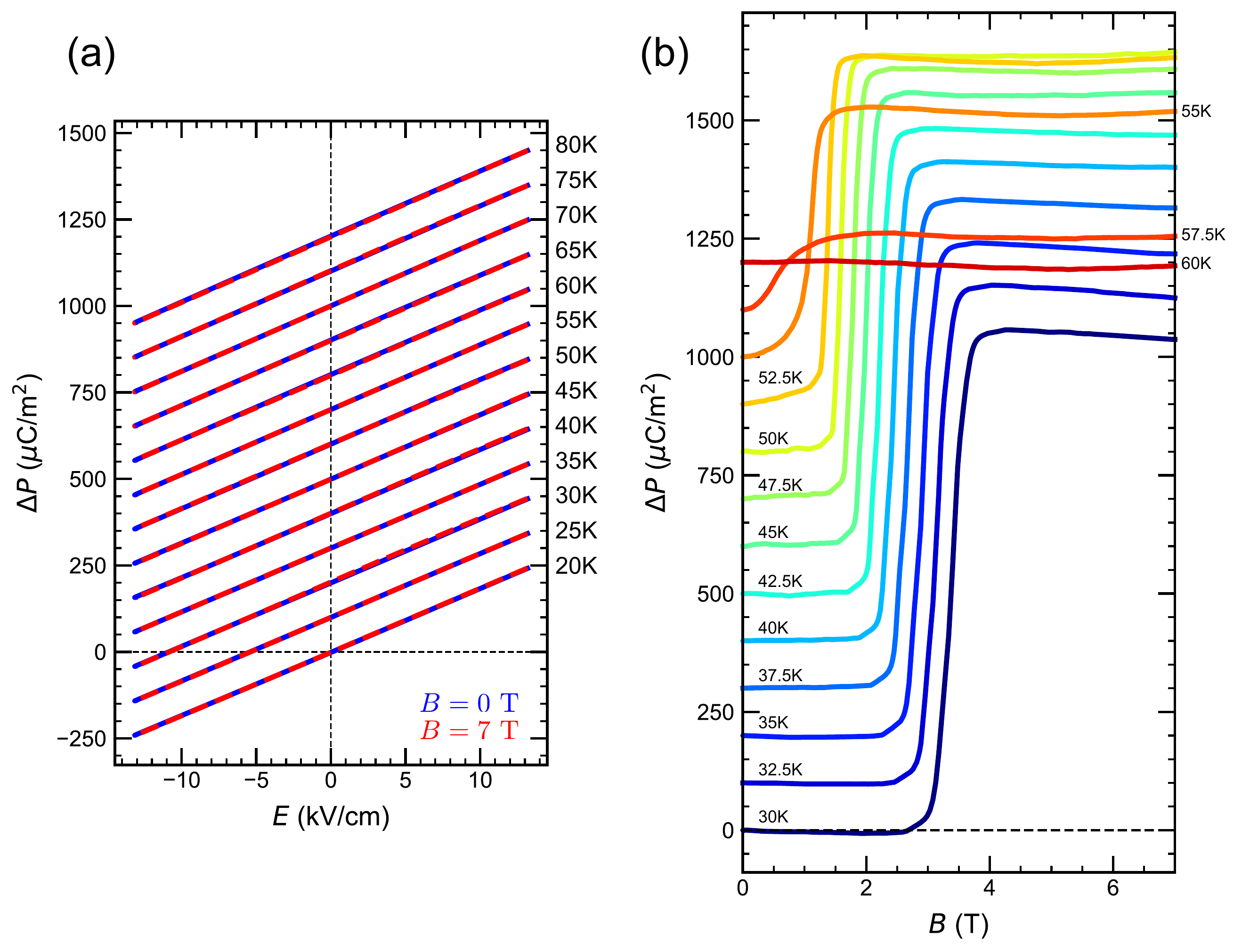}
  \caption{(a) $P$-$E$ curves and (b) $P$-$B$ curves of $\FZMO$ at selected temperatures. The $P$-$B$ curves were measured during the field-applying process. 
  Every $\Delta P$ is shifted by $\SI{100}{\mu C/m^2}$.}
  \label{FZMO_PE}
\end{figure}

The electric field dependence of the polarization is shown in Fig.~\ref{FZMO_PE}(a).
The polarization linearly changes with the electric field, and no phase transition is observed in the range of $|E| \leq \SI{13.3}{kV/cm}$.
The magnetic field dependence of the polarization is shown in Fig.~\ref{FZMO_PE}(b).
The magnetic-order-induced polarization $P_s$ gradually decreases as the temperature approaches $\SI{57.5}{K}$ and disappears above $\SI{60}{K}$.

\subsection*{Multifield contributions from the term related to the dielectric constant}
\begin{figure}[htb]
  \centering
  \includegraphics[width=6cm]{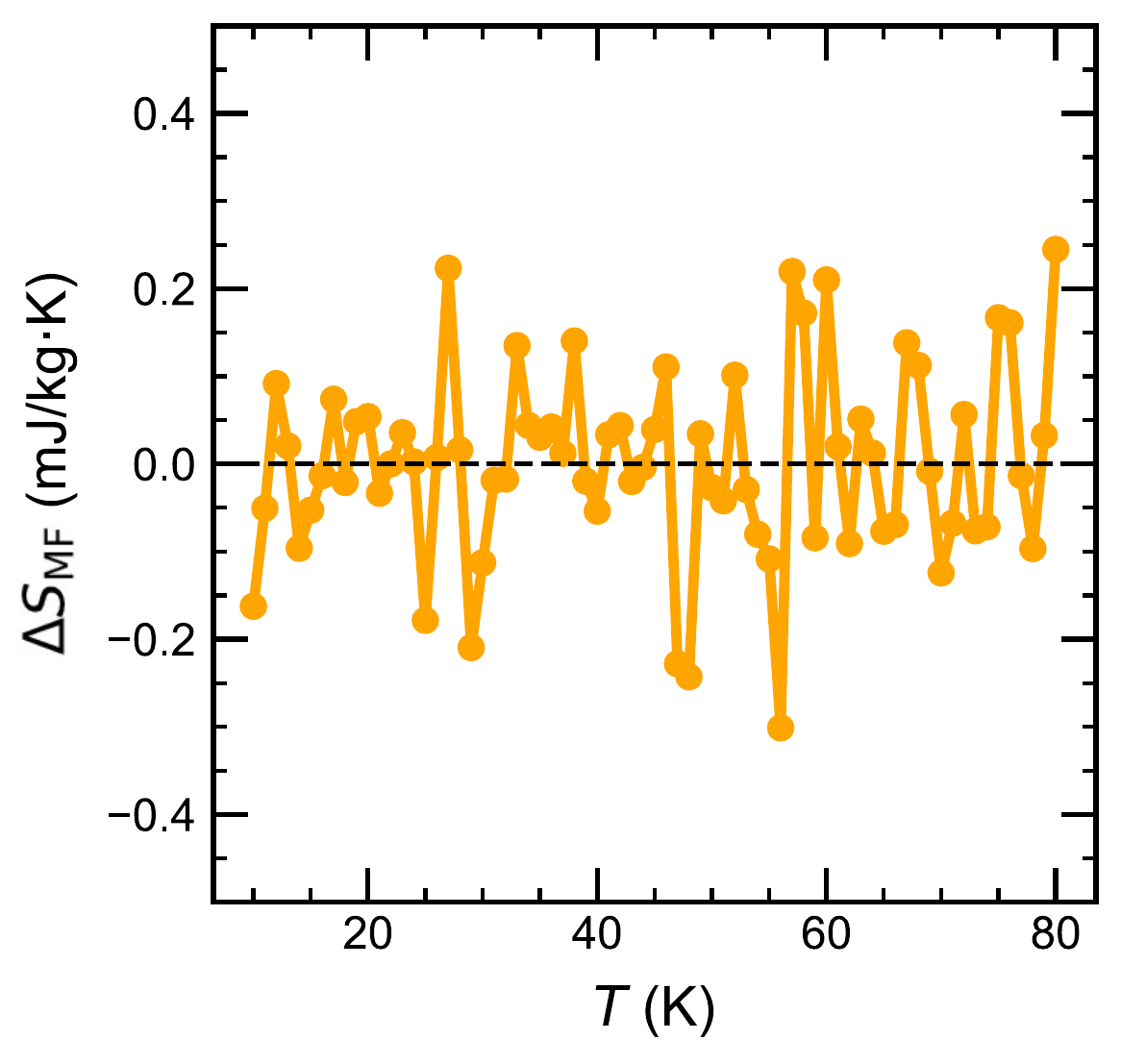}
  \caption{
  Temperature dependence of the second term in Eq.~(\ref{tag12}) for $\FZMO$ at $B_0 = \SI{7}{T}$ and $E_0 = \SI{10}{kV/cm}$.
  }
  \label{crossCE_PE}
\end{figure}

From Eq.~(4) in the main text, the multifield contribution to the MFCE, $\DeltaSME$, assuming that the electric field dependence of the polarization is linear up to $E=E_0$, is obtained:
\begin{equation}
      \DeltaSME 
      = \pdv{T} \Bigl(P_{E=0}(T,B_0) - P_{E=0}(T,0)\Bigl)E_0 + \frac{\epsilon_0}{2}\pdv{T}\Bigl( \chi^E(T,B_0) - \chi^E(T,0) \Bigl)E_0^2.
      % \tag{S6}
      \label{tag12}
\end{equation}
Figure~\ref{crossCE_PE} shows the second term in Eq.~(\ref{tag12}) for $\FZMO$.
The maximum value of the first term in Eq.~(\ref{tag12}) is $\SI{36}{mJ/kg\cdot K}$ (see Fig.~3(c) in the main text), indicating that the multifield contribution to the MFCE from the second term is negligible.  

\subsection*{Multifield contributions in various magnetoelectric multiferroics}
\begin{figure}[htb]
  \centering
  \includegraphics[width=\hsize]{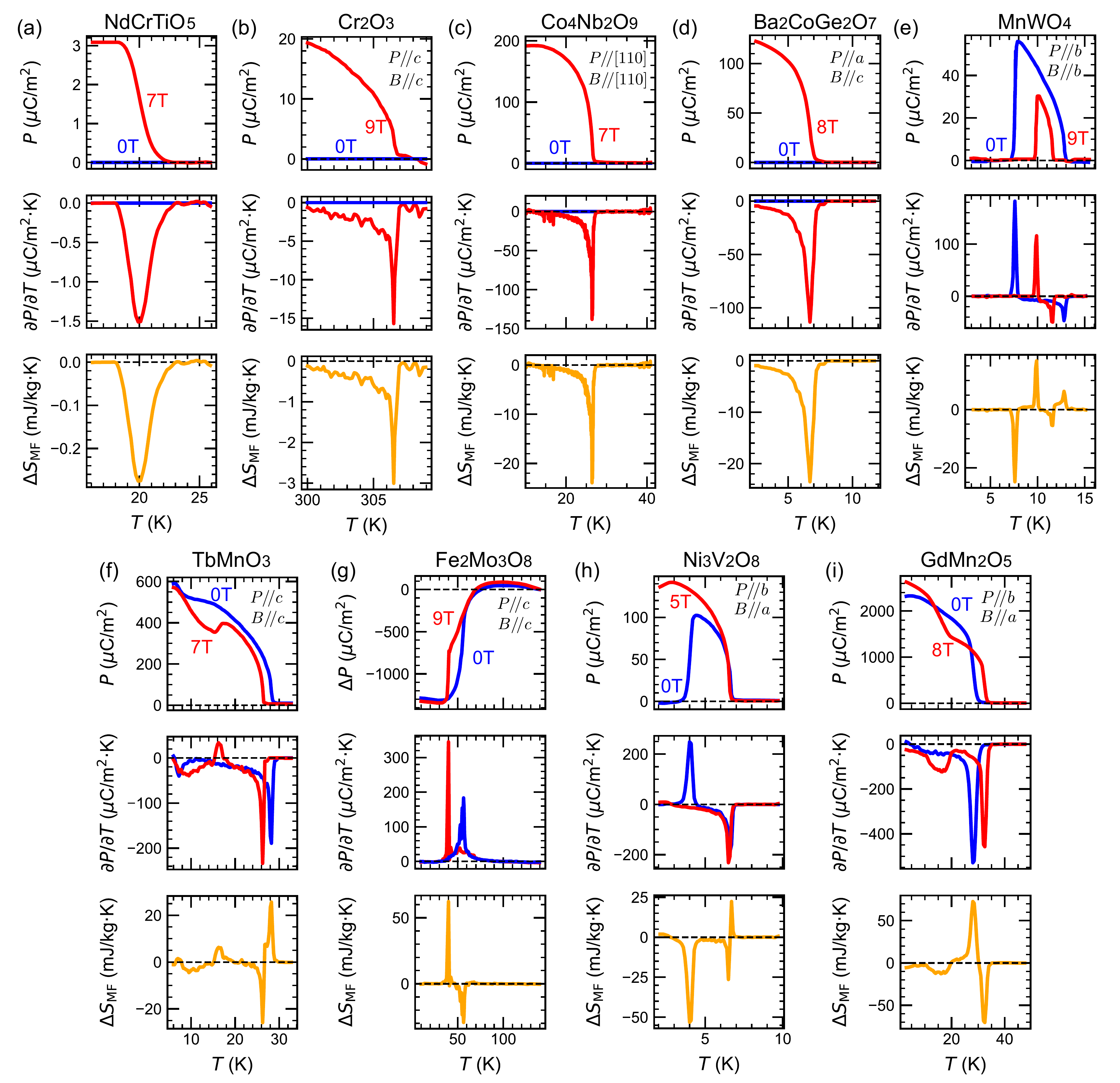}
  \caption{
  Temperature dependences of the polarization $P$ (top), $\partial P/\partial T$ (middle), and  $\DeltaSME$ (bottom) in  (a) \ce{NdCrTiO5}, (b) \ce{Cr2O3}, (c) \ce{Co4Nb2O9}, (d) \ce{Ba2CoGe2O7}, (e) \ce{MnWO4}, (f) \ce{TbMnO3}, (g) \ce{Fe2Mo3O8}, (h) \ce{Ni3V2O8}, and (i) \ce{GdMn2O5}.}
  \label{first_term}
\end{figure}

From Eqs.~(2) and (4) in the main text, the multifield contribution $\DeltaSME$ is given by the following equation;
\begin{equation}
\begin{aligned}
    \DeltaSME &= \int_0^{E_0}\left[\left(\pdv{P}{T}\right)_{B=B_0,E}-\left(\pdv{P}{T}\right)_{B=0,E}\right]\dd{E} \\
    &=  \pdv{T} \left[P_{E=0}(T,B_0) - P_{E=0}(T,0)\right]E_0  + \frac{\epsilon_0}{2}\pdv{T}\left[ \chi^E(T,B_0) - \chi^E(T,0) \right]E_0^2 \\
    & \simeq \pdv{T} \left[P_{E=0}(T,B_0) - P_{E=0}(T,0)\right]E_0, 
\end{aligned}
\end{equation}
where we assume that the electric field dependence of the polarization is linear up to $E_0 = \SI{10}{kV/cm}$ and that the contribution from the second $\chi^E$ term is negligible.
Figure~\ref{first_term} shows the magnitudes of $\DeltaSME$ for selected magnetoelectric compounds: \ce{NdCrTiO5} [42], \ce{Cr2O3} [43], \ce{Co4Nb2O9} [44], \ce{Ba2CoGe2O7} [45], \ce{MnWO4} [46], \ce{TbMnO3} [47], \ce{Fe2Mo3O8} [30], \ce{Ni3V2O8} [48], and \ce{GdMn2O5} [49].
The $\DeltaSME$ values for \ce{NdCrTiO5}, \ce{Co4Nb2O9}, and \ce{TbMnO3} are calculated for $B_0 = \SI{7}{T}$ and $E_0 = \SI{10}{kV/cm}$, which are the same conditions as those used for the estimation of $\DeltaSME$ in \ce{(Fe_{0.95}Zn_{0.05})2Mo3O8}.
The other materials are not calculated at $B_0 = \SI{7}{T}$ because $P$-$T$ curves at $\SI{7}{T}$ are not reported for these materials.
The $\DeltaSME$ values for \ce{Ba2CoGe2O7} and \ce{GdMn2O5} are calculated at $B_0 = \SI{8}{T}$ and $E_0 = \SI{10}{kV/cm}$.
The $\DeltaSME$ values for \ce{Cr2O3}, \ce{MnWO4}, and \ce{Fe2Mo3O8} are calculated at $B_0 = \SI{9}{T}$ and $E_0 = \SI{10}{kV/cm}$.
The $\DeltaSME$ values for \ce{Ni3V2O8} is calculated at $B_0 = \SI{5}{T}$ and $E_0 = \SI{10}{kV/cm}$.
Note that the sign of $\DeltaSME$ is determined by the difference of $\partial P/\partial T$ between $B=B_0$ and $B=0$, which is independent of the sign of magnetocaloric effect because $\DeltaSM$ is given by the temperature derivative of the magnetization $M$, i.e., $\partial M/\partial T$.

Finally, we comment on the adiabatic temperature change arising from the magnetoelectric cross-correlation.
In addition to the isothermal entropy change, the adiabatic temperature change $\Delta T_{\rm ad}$ is another key quantity characterizing the caloric effect.
Assuming that the specific heat $C_p$ is independent of the applied magnetic and electric fields, the adiabatic temperature change from the magnetoelectric cross-correlation, $\Delta T_{\rm MF}$, can be approximated as $\Delta T_{\rm MF} \simeq -(T/C_p) \DeltaSME$.
Table \ref{table2} summarizes $\Delta S_{\rm MF}$ and $\Delta T_{\rm MF}$ in various multiferroic compounds, where $\Delta T_{\rm MF}$ is estimated by using the specific heat $C_p$ at $B=0$ and at $E=0$.

\begin{table}[tb]
\caption{Multifield contribution to the MFCE in various multiferroics. $\Delta S_{\rm MF}$ is calculated at $B=B_0$ and $E_0=\SI{10}{kV/cm}$, and the maximum value is shown.
$\Delta T_{\rm MF}$ represents the estimated adiabatic temperature change contributed from the magnetoelectric cross-correlation, based on $\Delta T_{\rm MF}=-(T/C_p)\Delta S_{\rm MF}$.
``Linear ME effect'' and ``Magnetic-order'' in the last column represent the linear magnetoelectric effect and magnetic-order-induced polarization, respectively.
All data are from single crystals, except for the polarization and specific heat data of polycrystalline NdCrTiO$_5$ [42] , and the specific heat data of polycrystalline Co$_4$Nb$_2$O$_9$ [56]. }

\label{table2}
\begin{ruledtabular}
\begin{tabular}{lcccccc}
& $T$ & $B_0$  & $\abs{\Delta S_{\rm MF}}$ & $C_p$ & $\abs{\Delta T_{\rm MF}}$  & Main mechanism \\[0.5ex]
 & ($\SI{}{K}$) & ($\SI{}{T}$)  & ($\SI{}{mJ/kg\cdot K}$) & ($\SI{}{J/kg \cdot K}$) & ($\SI{}{mK}$) & of polarization \\[0.5ex]
\hline
\ce{(Fe_{0.95}Zn_{0.05})2Mo3O8} [29] &  $56$ & $7$ & $36$ & $130$ & $16$ & Magnetic-order \\
\ce{NdCrTiO5} [42] & $20$ & $7$ & $0.28$ & $26$ & $0.21$ & Linear ME effect \\
\ce{Cr2O3} [43, 55] & $307$ & $9$ & $3.0$ & $931$ & $0.99$ & Linear ME effect \\
\ce{Co4Nb2O9} [44, 56] & $26$ & $7$ & $24$ & $90$ & $7.0$ &  Linear ME effect \\
\ce{Ba2CoGe2O7} [45] & $6.7$ & $8$ & $24$ & $17$ & $9.2$ &  Magnetic-order \\
\ce{MnWO4} [46] &$7.6$ & $9$ & $25$ & $29$ & $6.5$ & Magnetic-order \\
\ce{TbMnO3} [47, 57] & $26$ & $7$ & $26$ & $48$ & $15$ & Magnetic-order \\
\ce{Ni3V2O8} [48, 58] & $4.0$ & $5$ & $53$ & $35$ & $6.1$ & Magnetic-order \\
\ce{Fe2Mo3O8} [30, 36]  & $40$ & $9$ & $62$ & $52$ & $48$ & Magnetic-order \\
\ce{GdMn2O5} [49, 59]  & $28$ & $8$ & $73$ & $80$ & $25$  & Magnetic-order \\
\end{tabular}
\end{ruledtabular}
\end{table}

\clearpage

\bibliographystyle{aip}
\bibliography{ref}

\begin{thebibliography}{10}

\bibitem{kimura2003magnetic}
T.~Kimura, T.~Goto, H.~Shintani, K.~Ishizaka, T.-h. Arima, and Y.~Tokura,
\newblock Magnetic control of ferroelectric polarization,
\newblock Nature {\bf 426}, 55 (2003).

\bibitem{tokura2014multiferroics}
Y.~Tokura, S.~Seki, and N.~Nagaosa,
\newblock Multiferroics of spin origin,
\newblock Rep. Prog. Phys. {\bf 77}, 076501 (2014).

\bibitem{fiebig2016evolution}
M.~Fiebig, T.~Lottermoser, D.~Meier, and M.~Trassin,
\newblock The evolution of multiferroics,
\newblock Nat. Rev. Mater. {\bf 1}, 1 (2016).

\bibitem{ideue2017giant}
T.~Ideue, T.~Kurumaji, S.~Ishiwata, and Y.~Tokura,
\newblock Giant thermal hall effect in multiferroics,
\newblock Nat. Mater. {\bf 16}, 797 (2017).

\bibitem{kim2024thermal}
H.-L. Kim, T.~Saito, H.~Yang, H.~Ishizuka, M.~J. Coak, J.~H. Lee, H.~Sim, Y.~S. Oh, N.~Nagaosa, and J.-G. Park,
\newblock Thermal hall effects due to topological spin fluctuations in \ce{YMnO3},
\newblock Nat. Commun. {\bf 15}, 243 (2024).

\bibitem{hirokane2020nonreciprocal}
Y.~Hirokane, Y.~Nii, H.~Masuda, and Y.~Onose,
\newblock Nonreciprocal thermal transport in a multiferroic helimagnet,
\newblock Sci. Adv. {\bf 6}, eabd3703 (2020).

\bibitem{takagi2016thermal}
R.~Takagi, Y.~Tokunaga, T.~Ideue, Y.~Taguchi, Y.~Tokura, and S.~Seki,
\newblock Thermal generation of spin current in a multiferroic helimagnet,
\newblock APL Mater. {\bf 4}, 032502 (2016).

\bibitem{seki2015thermal}
S.~Seki, T.~Ideue, M.~Kubota, Y.~Kozuka, R.~Takagi, M.~Nakamura, Y.~Kaneko, M.~Kawasaki, and Y.~Tokura,
\newblock Thermal generation of spin current in an antiferromagnet,
\newblock Phys. Rev. Lett. {\bf 115}, 266601 (2015).

\bibitem{ikeda2023magnetoelectrocaloric}
R.~Ikeda, T.~Kurumaji, Y.~Tokunaga, and T.-h. Arima,
\newblock Magnetoelectric coupling based caloric effect in multiferroic \ce{GdFeO3},
\newblock J. Phys. Soc. Jpn. {\bf 93}, 094709 (2024).

\bibitem{moya2014caloric}
X.~Moya, S.~Kar-Narayan, and N.~D. Mathur,
\newblock Caloric materials near ferroic phase transitions,
\newblock Nat. Mater. {\bf 13}, 439 (2014).

\bibitem{hou2022materials}
H.~Hou, S.~Qian, and I.~Takeuchi,
\newblock Materials, physics and systems for multicaloric cooling,
\newblock Nat. Rev. Mater. {\bf 7}, 633 (2022).

\bibitem{tishin2016magnetocaloric}
A.~M. Tishin and Y.~I. Spichkin,
\newblock The magnetocaloric effect and its applications,
\newblock CRC Press, 2003.

\bibitem{liu2016direct}
Y.~Liu, J.~F. Scott, and B.~Dkhil,
\newblock Direct and indirect measurements on electrocaloric effect: Recent developments and perspectives,
\newblock Appl. Phys. Rev. {\bf 3}, 031102 (2016).

\bibitem{cong2019colossal}
D.~Cong, W.~Xiong, A.~Planes, Y.~Ren, L.~Ma{\~n}osa, P.~Cao, Z.~Nie, X.~Sun, Z.~Yang, X.~Hong, et~al.,
\newblock Colossal elastocaloric effect in ferroelastic \ce{Ni}-\ce{Mn}-\ce{Ti} alloys,
\newblock Phys. Rev. Lett. {\bf 122}, 255703 (2019).

\bibitem{matsunami2015giant}
D.~Matsunami, A.~Fujita, K.~Takenaka, and M.~Kano,
\newblock Giant barocaloric effect enhanced by the frustration of the antiferromagnetic phase in \ce{Mn3GaN},
\newblock Nat. Mater. {\bf 14}, 73 (2015).

\bibitem{starkov2014multicaloric}
A.~Starkov and I.~Starkov,
\newblock Multicaloric effect in a solid: new aspects,
\newblock J. Exp. Theor. Phys. {\bf 119}, 258 (2014).

\bibitem{starkov2016generalized}
I.~A. Starkov and A.~S. Starkov,
\newblock A generalized thermodynamic theory of the multicaloric effect in single-phase solids,
\newblock Int. J. Solids Struct. {\bf 100}, 187 (2016).

\bibitem{stern2018multicaloric}
E.~Stern-Taulats, T.~Cast{\'a}n, L.~Ma{\~n}osa, A.~Planes, N.~D. Mathur, and X.~Moya,
\newblock Multicaloric materials and effects,
\newblock MRS Bull. {\bf 43}, 295 (2018).

\bibitem{kosugi2021giant}
Y.~Kosugi, M.~Goto, Z.~Tan, D.~Kan, M.~Isobe, K.~Yoshii, M.~Mizumaki, A.~Fujita, H.~Takagi, and Y.~Shimakawa,
\newblock Giant multiple caloric effects in charge transition ferrimagnet,
\newblock Sci. Rep. {\bf 11}, 12682 (2021).

\bibitem{qian2022multicaloric}
H.~Qian, J.~Guo, Z.~Wei, and J.~Liu,
\newblock Multicaloric effect in synergic magnetostructural phase transformation \ce{Ni}-\ce{Mn}-\ce{Ga}-\ce{In} alloys,
\newblock Phy. Rev. Mater. {\bf 6}, 054401 (2022).

\bibitem{manosa2023cross}
L.~Ma{\~n}osa, E.~Stern-Taulats, A.~Gr{\`a}cia-Condal, and A.~Planes,
\newblock Cross-coupling contribution to the isothermal entropy change in multicaloric materials,
\newblock J. Phys.: Energy {\bf 5}, 024016 (2023).

\bibitem{liu2016large}
Y.~Liu, L.~C. Phillips, R.~Mattana, M.~Bibes, A.~Barth{\'e}l{\'e}my, and B.~Dkhil,
\newblock Large reversible caloric effect in \ce{FeRh} thin films via a dual-stimulus multicaloric cycle,
\newblock Nat. Commun. {\bf 7}, 11614 (2016).

\bibitem{vopson2012multicaloric}
M.~M. Vopson,
\newblock The multicaloric effect in multiferroic materials,
\newblock Solid State Commun. {\bf 152}, 2067 (2012).

\bibitem{vopson2013theory}
M.~M. Vopson,
\newblock Theory of giant-caloric effects in multiferroic materials,
\newblock J. Phys. D: Appl. Phys. {\bf 46}, 345304 (2013).

\bibitem{VOPSON201614}
M.~M. Vopson,
\newblock The induced magnetic and electric fields’ paradox leading to multicaloric effects in multiferroics,
\newblock Solid State Commun. {\bf 231-232}, 14 (2016).

\bibitem{edstrom2020prediction}
A.~Edstr{\"o}m and C.~Ederer,
\newblock Prediction of a giant magnetoelectric cross-caloric effect around a tetracritical point in multiferroic \ce{SrMnO3},
\newblock Phys. Rev. Lett. {\bf 124}, 167201 (2020).

\bibitem{planes2014thermodynamics}
A.~Planes, T.~Castan, and A.~Saxena,
\newblock Thermodynamics of multicaloric effects in multiferroics,
\newblock Philos. Mag. {\bf 94}, 1893 (2014).

\bibitem{li2024cross}
S.~Li, N.~Fan, L.~Bellaiche, and B.~Xu,
\newblock Cross-caloric effect in multiferroic bismuth ferrite,
\newblock Phys. Rev. B {\bf 109}, 214110 (2024).

\bibitem{kurumaji2015doping}
T.~Kurumaji, S.~Ishiwata, and Y.~Tokura,
\newblock Doping-tunable ferrimagnetic phase with large linear magnetoelectric effect in a polar magnet \ce{Fe2Mo3O8},
\newblock Phys. Rev. X {\bf 5}, 031034 (2015).

\bibitem{mccarroll1957some}
W.~H. McCarroll, L.~Katz, and R.~Ward,
\newblock Some ternary oxides of tetravalent molybdenum1, 2,
\newblock J. Am. Chem. Soc. {\bf 79}, 5410 (1957).

\bibitem{cotton1964metal}
F.~A. Cotton,
\newblock Metal atom clusters in oxide systems,
\newblock Inorg. Chem. {\bf 3}, 1217 (1964).

\bibitem{FZMO_prefer}
F.~Varret, H.~Czeskleba, F.~Hartmann-Boutron, and P.~Imbert,
\newblock {{\'E}tude par effet M{\"o}ssbauer de l'ion \ce{Fe^2+} en sym{\'e}trie trigonale dans les compos{\'e}s du type \ce{(Fe,M)2Mo3O8} (M = Mg, Zn, Mn, Co, Ni) et propri{\'e}t{\'e}s magn{\'e}tiques de \ce{(Fe,Zn)2Mo3O8}},
\newblock Inorg. Chem. {\bf 33}, 549 (1972).

\bibitem{bertrand1975structural}
D.~Bertrand and H.~Kerner-Czeskleba,
\newblock Structural and magnetic study of iron-group molybdates,
\newblock J. Phys. (Paris) {\bf 36}, 379 (1975).

\bibitem{reschke2020structure}
S.~Reschke, A.~A. Tsirlin, N.~Khan, L.~Prodan, V.~Tsurkan, I.~Kezsmarki, and J.~Deisenhofer,
\newblock Structure, phonons, and orbital degrees of freedom in \ce{Fe2Mo3O8},
\newblock Phys. Rev. B {\bf 102}, 094307 (2020).

\bibitem{matsuura2023thermodynamic}
K.~Matsuura, Y.~Nishizawa, M.~Kriener, T.~Kurumaji, H.~Oike, Y.~Tokura, and F.~Kagawa,
\newblock Thermodynamic determination of the equilibrium first-order phase-transition line hidden by hysteresis in a phase diagram,
\newblock Sci. Rep. {\bf 13}, 6876 (2023).

\bibitem{wang2015unveiling}
Y.~Wang, G.~L. Pascut, B.~Gao, T.~A. Tyson, K.~Haule, V.~Kiryukhin, and S.-W. Cheong,
\newblock Unveiling hidden ferrimagnetism and giant magnetoelectricity in polar magnet \ce{Fe2Mo3O8},
\newblock Sci. Rep. {\bf 5}, 12268 (2015).

\bibitem{yonezawa2013first}
S.~Yonezawa, T.~Kajikawa, and Y.~Maeno,
\newblock First-order superconducting transition of {S}r$_2${R}u{O}$_4$,
\newblock Phys. Rev. Lett. {\bf 110}, 077003 (2013).

\bibitem{krenke2005inverse}
T.~Krenke, E.~Duman, M.~Acet, E.~F. Wassermann, X.~Moya, L.~Ma{\~n}osa, and A.~Planes,
\newblock Inverse magnetocaloric effect in ferromagnetic \ce{Ni}-\ce{Mn}-\ce{Sn} alloys,
\newblock Nat. Mater. {\bf 4}, 450 (2005).

\bibitem{tran2022direct}
H.~B. Tran, T.~Fukushima, H.~Momida, K.~Sato, Y.~Makino, and T.~Oguchi,
\newblock Direct and inverse magnetocaloric effects in \ce{FeRh} alloy,
\newblock J. Alloys Compd. {\bf 926}, 166718 (2022).

\bibitem{citation-key}
See Supplemental Material for the results of the polarization measurements, the multi-field contribution $\Delta S_{\rm MF}$ from the dielectric properties in $\FZMO$, and $\Delta S_{\rm MF}$ in various multiferroics. The Supplemental Material additionally includes Refs.~\cite{murtazaev2001heat,xie2023colossal, o2014magnetic, zhang2008specific,zheng2019abnormal}.

\bibitem{kurumaji2017diagonal}
T.~Kurumaji, S.~Ishiwata, and Y.~Tokura,
\newblock Diagonal magnetoelectric susceptibility and effect of fe doping in the polar ferrimagnet \ce{Mn2Mo3O8},
\newblock Phys. Rev. B {\bf 95}, 045142 (2017).

\bibitem{PhysRevB.85.024415}
J.~Hwang, E.~S. Choi, H.~D. Zhou, J.~Lu, and P.~Schlottmann,
\newblock Magnetoelectric effect in \ce{NdCrTiO5},
\newblock Phys. Rev. B {\bf 85}, 024415 (2012).

\bibitem{PhysRevB.87.180408}
A.~Iyama and T.~Kimura,
\newblock Magnetoelectric hysteresis loops in \ce{Cr2O3} at room temperature,
\newblock Phys. Rev. B {\bf 87}, 180408 (2013).

\bibitem{khanh2016magnetoelectric}
N.~D. Khanh, N.~Abe, H.~Sagayama, A.~Nakao, T.~Hanashima, R.~Kiyanagi, Y.~Tokunaga, and T.~Arima,
\newblock Magnetoelectric coupling in the honeycomb antiferromagnet \ce{Co4Nb2O9},
\newblock Phys. Rev. B {\bf 93}, 075117 (2016).

\bibitem{yi2008multiferroicity}
H.~Yi, Y.~Choi, S.~Lee, and S.-W. Cheong,
\newblock Multiferroicity in the square-lattice antiferromagnet of \ce{Ba2CoGe2O7},
\newblock Appl. Phys. Lett. {\bf 92}, 212904 (2008).

\bibitem{PhysRevB.74.184431}
A.~H. Arkenbout, T.~T.~M. Palstra, T.~Siegrist, and T.~Kimura,
\newblock Ferroelectricity in the cycloidal spiral magnetic phase of \ce{MnWO4},
\newblock Phys. Rev. B {\bf 74}, 184431 (2006).

\bibitem{kimura2005magnetoelectric}
T.~Kimura, G.~Lawes, T.~Goto, Y.~Tokura, and A.~P. Ramirez,
\newblock Magnetoelectric phase diagrams of orthorhombic \ce{RMnO3} (\ce{R=Gd}, \ce{Tb}, and \ce{Dy}),
\newblock Phys. Rev. B {\bf 71}, 224425 (2005).

\bibitem{PhysRevLett.95.087205}
G.~Lawes, A.~B. Harris, T.~Kimura, N.~Rogado, R.~J. Cava, A.~Aharony, O.~Entin-Wohlman, T.~Yildirim, M.~Kenzelmann, C.~Broholm, and A.~P. Ramirez,
\newblock Magnetically driven ferroelectric order in \ce{Ni3V2O8},
\newblock Phys. Rev. Lett. {\bf 95}, 087205 (2005).

\bibitem{PhysRevB.94.174446}
S.~H. Bukhari, T.~Kain, M.~Schiebl, A.~Shuvaev, A.~Pimenov, A.~M. Kuzmenko, X.~Wang, S.-W. Cheong, J.~Ahmad, and A.~Pimenov,
\newblock Magnetoelectric phase diagrams of multiferroic \ce{GdMn2O5},
\newblock Phys. Rev. B {\bf 94}, 174446 (2016).

\bibitem{andrade2019multicaloric}
V.~Andrade, A.~Amirov, D.~Yusupov, B.~Pimentel, N.~Barroca, A.~Pires, J.~Belo, A.~Pereira, M.~Valente, J.~Ara{\'u}jo, et~al.,
\newblock Multicaloric effect in a multiferroic composite of \ce{Gd5(Si,Ge)4} microparticles embedded into a ferroelectric pvdf matrix,
\newblock Sci. Rep. {\bf 9}, 18308 (2019).

\bibitem{10.1063/1.5090599}
F.-X. Liang, J.-Z. Hao, F.-R. Shen, H.-B. Zhou, J.~Wang, F.-X. Hu, J.~He, J.-R. Sun, and B.-G. Shen,
\newblock Experimental study on coupled caloric effect driven by dual fields in metamagnetic heusler alloy \ce{Ni50Mn35In15},
\newblock APL Mater. {\bf 7}, 051102 (2019).

\bibitem{stern2017giant}
E.~Stern-Taulats, T.~Cast{\'a}n, A.~Planes, L.~H. Lewis, R.~Barua, S.~Pramanick, S.~Majumdar, and L.~Ma{\~n}osa,
\newblock Giant multicaloric response of bulk \ce{Fe49Rh51},
\newblock Phys. Rev. B {\bf 95}, 104424 (2017).

\bibitem{starkov2015rebuttalthemulticaloriceffect}
I.~A. Starkov and A.~S. Starkov,
\newblock Rebuttal of "the multicaloric effect in multiferroic materials",
\newblock arXiv:1602.04238  (2015).

\bibitem{MENG2013567}
H.~Meng, B.~Li, W.~Ren, and Z.~Zhang,
\newblock Coupled caloric effects in multiferroics,
\newblock Phys. Lett. A {\bf 377}, 567 (2013).

\bibitem{murtazaev2001heat}
A.~Murtazaev, S.~B. Abdulvagidov, A.~Aliev, and O.~Musaev,
\newblock Heat capacity of a \ce{Cr2O3} antiferromagnet near the critical temperature,
\newblock Phys. Solid State {\bf 43}, 1103 (2001).

\bibitem{xie2023colossal}
Y.~Xie, F.~Guo, H.~Li, B.~Tao, and H.~Chang,
\newblock Colossal magneto-nonlinear-dielectric effect in magnetoelectric antiferromagnet tetracobalt diniobate,
\newblock J. Magn. Magn. Mater. {\bf 587}, 171372 (2023).

\bibitem{o2014magnetic}
D.~O’Flynn, M.~R. Lees, and G.~Balakrishnan,
\newblock Magnetic susceptibility and heat capacity measurements of single crystal \ce{TbMnO3},
\newblock J. Phys.: Condens. Matter {\bf 26}, 256002 (2014).

\bibitem{zhang2008specific}
Q.~Zhang, W.~Knafo, K.~Grube, H.~V. L{\"o}hneysen, C.~Meingast, and T.~Wolf,
\newblock Specific heat of the kagom{\'e} mixed compounds ({C}o$_{1-x}${N}i$_x$)$_3${V}$_2${O}$_8$,
\newblock Phys. B: Condens. Matter {\bf 403}, 1404 (2008).

\bibitem{zheng2019abnormal}
S.~Zheng, J.~Gong, Y.~Li, C.~Li, Y.~Tang, J.~Zhang, L.~Lin, Z.~Yan, X.~Jiang, S.~Cheong, et~al.,
\newblock Abnormal dependence of multiferroicity on high-temperature electro-poling in \ce{GdMn2O5},
\newblock J. Appl. Phys. {\bf 126}, 174104 (2019).

\end{thebibliography}

\end{document}